\documentclass{IEEEtran}
\usepackage{graphicx}
\usepackage{bm}
\usepackage{cite}
\usepackage{pgfplots}
\usepackage{color}
\usepackage{algorithm}
\usepackage{amsmath,amssymb,amsthm,dsfont,mathrsfs}

\def\diag{{\mathrm{diag}}}
\def\tr{{\mathrm{tr}}}
\def\A{{\mathbf{A}}}
\def\Y{{\mathbf{Y}}}
\def\S{{\mathbf{S}}}

\def\I{{\mathbf{I}}}
\def\N{{\mathbf{N}}}

\def\y{{\mathbf{y}}}
\def\s{{\mathbf{s}}}
\def\a{{\mathbf{a}}}
\def\n{{\mathbf{n}}}

\def\H{{\mathbf{H}}}
\def\X{{\mathbf{X}}}
\def\Y{{\mathbf{Y}}}
\def\h{{\mathbf{h}}}

\def\F{{\mathbf{F}}}
\def\v{{\mathbf{v}}}

\def\c{{\mathbf{c}}}

\def\w{{\mathbf{w}}}

\def\f{{\mathbf{f}}}
\def\e{{\mathbf{e}}}
\def\E{{\mathbf{E}}}

\def\t{{\mathbf{t}}}

\begin{document}
%
\title{FDD Massive MIMO Channel Estimation with Arbitrary 2D-Array Geometry}

\author{{Jisheng~Dai},
        {An~Liu},
        and~{Vincent K. N.  Lau}


\thanks{J. Dai is with the Department of Electronic Engineering, Jiangsu University, Zhenjiang 212013, China, and also with the Department of ECE, The Hong Kong University of Science and Technology, Hong Kong (e-mail: jsdai@ujs.edu.cn).}
\thanks{A. Liu was with the Department of ECE, The Hong Kong University of Science and Technology. He is now with the College of Information Science and Electronic Engineering, Zhejiang University (e-mail: wendaolstr@gmail.com).}
\thanks{ V. Lau is  with the Department of ECE, The Hong Kong University of Science and Technology, Hong Kong (e-mail: eeknlau@ust.hk). }
}

\maketitle

\begin{abstract}
This paper addresses the problem of downlink channel estimation in frequency-division duplexing (FDD) massive multiple-input multiple-output (MIMO) systems.
The existing methods usually exploit hidden sparsity  under a discrete Fourier transform (DFT) basis to estimate the downlink channel. However, there are at least two shortcomings of these DFT-based methods: 1) they  are applicable to uniform linear arrays (ULAs) only, since the DFT basis requires a special structure of ULAs; and 2) they always suffer from a performance loss due to the leakage of energy over some DFT bins. To deal with the above shortcomings, we introduce an off-grid model
 for downlink channel sparse representation with arbitrary 2D-array antenna geometry, and propose an efficient sparse Bayesian learning (SBL) approach for the sparse channel recovery and off-grid refinement.
The main idea of the proposed off-grid method is to consider the sampled grid points  as adjustable parameters.
Utilizing  an in-exact block majorization-minimization (MM) algorithm,  the grid points are refined iteratively to minimize the off-grid gap.
Finally, we further extend the solution to uplink-aided  channel estimation by exploiting the angular reciprocity between downlink and uplink channels, which brings enhanced recovery performance.
\end{abstract}


\IEEEpeerreviewmaketitle

\section{Introduction}

Massive multiple-input multiple-output (MIMO)  has attracted significant attention in wireless communications, and has been widely considered as a key candidate technology to meet the capacity demand in  5G wireless networks \cite{marzetta2010noncooperative,rusek2013scaling}. To fully reap the benefit of excessive base station ($\mathrm{BS}$) antennas, knowledge of channel state information at the transmitter (CSIT) is essentially required \cite{shen2017high}. Many research efforts have  been devoted to time-division duplexing (TDD) massive MIMO, because the CSIT in the TDD mode can be obtained by  exploiting channel reciprocity, where the pilot-aided training overhead is proportional to the number of active mobile users (MUs) only \cite{larsson2014massive,hoydis2013massive}. However,
in the frequency-division duplexing (FDD) mode, the conventional  training overhead for the CSIT acquisition grows proportionally with the BS antenna size \cite{hassibi2003much,lu2014overview}, which can be quite large in massive MIMO systems. Hence, it appears to be an extremely challenging task to obtain accurate CSIT in FDD massive MIMO systems.

Fortunately, due to the limited local scattering effect in the propagation environment, the elements in the massive MIMO channel are highly correlated.  Many works have shown that the effective dimension of a massive MIMO channel is much less than its original dimension \cite{gao2015spatially,hoydis2012channel,rao2014distributed,gao2015priori}. Specifically, if the BS is equipped with a large uniform linear array (ULA),  the massive MIMO channel  has an $approximately$ sparse representation under the discrete Fourier transform (DFT) basis \cite{chen2016pilot,rao2014distributed,shen2016compressed}.
Exploiting such hidden sparsity,
many efficient downlink channel estimation and feedback algorithms have been proposed in recent years \cite{gao2015spatially,rao2015compressive,rao2014distributed,choi2014downlink,you2016channel,gao2016structured,gao2015priori,liu2016exploiting,liu2017closed}.
Nevertheless, it is worth noting that the validity of the DFT basis as a sparse representation of a massive MIMO channel depends on  ULAs.
When the antenna geometry deviates from a ULA, the aforementioned methods  will fail to work.

DFT-based channel estimation methods always have a performance loss, even for ULA systems, because of the leakage of energy in the DFT basis.
As shown in \cite{ding2015channel,ding2015compressed,ding2016dictionary}, the DFT basis actually provides a fixed sampling grid that discretely covers the angular domain of the massive MIMO system.  Since signals usually come from random directions,  the leakage energy caused by direction mismatch is unavoidable.
To achieve a better sparse representation, Ding and Rao \cite{ding2015channel,ding2015compressed,ding2016dictionary} considered an overcomplete DFT basis, which corresponds to a denser sampling grid on the angular domain.
The overcomplete DFT basis may still lead to a high direction mismatch, if the grid is not sufficiently dense.
On the other hand, if a very dense sampling grid is used,
the $l_1$-norm-based recovery methods may not work well due to high correlation between the basis vectors.
To overcome the leakage issue and to generalize for general antenna geometry, dictionary learning techniques were also proposed in \cite{ding2015channel,ding2015compressed,ding2016dictionary}. However, the standard dictionary learning approach has several drawbacks: 1)  its convergence is not theoretically guaranteed; and 2) learning a comprehensive dictionary requires collecting a large amount of channel measurements as training samples from all locations in a specific cell, which may pose great challenges in practical implementations.

In this paper, we consider a generic off-grid model for channel sparse representation of massive MIMO systems with an arbitrary 2D-array geometry, and we propose an efficient sparse Bayesian learning (SBL) approach \cite{tipping2001sparse,ji2008bayesian} for joint sparse channel recovery and off-grid refinement. The main idea of the proposed method is to consider the sampled grid points  as adjustable parameters. Then, we utilize an in-exact block majorization-minimization (MM) algorithm \cite{razaviyayn2014successive,sun2017majorization} to refine the grid points iteratively. After several iterations, the refined points will approach the actual directions of arrival/departure, so the proposed method can significantly alleviate  direction mismatch in the angular domain. The following summarizes the contributions of this paper.

\begin{itemize}
  \item {\bf Model-based Off-Grid Sparse Basis}

  We provide a novel off-grid model for massive MIMO channel sparse representation  with an arbitrary 2D-array geometry.
  Off-grid models have been applied widely to the direction-of-arrival in array signal processing \cite{yang2013off,liu2017off,dai2018sparse}.  However, the commonly used linear approximation off-grid model does not work well, especially when the grid is not sufficiently fine \cite{dai2017root}.  Our proposed model avoids using any approximations, and thus can significantly alleviate the modeling error.

\item {\bf Joint Sparse Channel Recovery and Off-Grid Refinement with Autonomous Learning}

We propose an SBL-based framework  based on in-exact block MM algorithm for joint sparse channel recovery and off-grid refinement.
The proposed solution outperforms $l_1$-norm recovery \cite{Donoho2008,donoho2006compressed,candes2006robust},\footnote{SBL methods include  $l_1$-norm-based methods as a special case when a maximum $a$ $posteriori$ (MAP) optimal estimate is
adopted with a fixed Laplace signal prior, and theoretical and empirical
results show that SBL methods with better priors can achieve enhanced
performance over  the $l_1$-norm-based methods \cite{ji2008bayesian,wipf2004sparse}.} and has an inherent  learning capability, so no prior knowledge about the sparsity level, noise variance or direction mismatch is required. We show that the solution converges to the stationary solution of the  optimization problem. Simulation results reveal substantial performance gains over the existing state-of-the-art baselines.

\item {\bf Enhanced Recovery Performance with Angular Reciprocity}

We further extend the solution to uplink-aided  channel estimation by exploiting angular reciprocity\footnote{We consider an FDD system, so the reciprocity of the channel realization between the uplink and downlink  does not hold. However, the directions of arrival and departure of the uplink are reciprocal with those of the downlink due to the fact that both the uplink and downlink face the same scattering structure \cite{ding2016dictionary}.} between downlink and uplink channels.
Characterizing the joint sparse structure with angular reciprocity was first addressed in \cite{ding2016dictionary}. However, it always has a performance loss due to the fact that the joint sparse structure only holds approximately. Our new extension  strictly
characterizes  the joint sparse structure by the inherent mechanism of the off-grid model, bringing enhanced recovery performance.

\end{itemize}

The rest of the paper is organized as follows. In Section II, we present the system model and review the state-of-the-art DFT-based channel estimation for  massive MIMO systems.
In Section III, we provide the SBL-based off-grid  channel estimation method for a linear array, and then, in Section IV, we extend it to an  arbitrary 2D-array geometry.
In Section V, we exploit angular reciprocity to improve channel estimation performance. Numerical experiments and
discussions  follow in Sections VI and VII, respectively.

$Notations:$ $\mathbb{C}$ denotes complex number, $\|\cdot\|_p$ denotes $p$-norm, $(\cdot)^T$ denotes transpose, $(\cdot)^H$ denotes Hermitian transpose,
$(\cdot)^\dag$ denotes pseudoinverse,
$\I$ denotes identity matrix,
$\A_\Omega$ denotes the sub-matrix formed by collecting the columns from $\Omega$,
$\mathcal{CN}(\cdot, \bm\mu, \bm\Sigma)$ denotes complex Gaussian distribution with mean $\bm\mu$ and variance $\bm\Sigma$, $\mathrm{supp}(\cdot)$ denotes the set of indices of nonzero elements, $\tr(\cdot)$ denotes trace operator, $\diag(\cdot)$  denotes diagonal operator, and $\odot$ denotes Hadamard product.

\section{Massive MIMO Channel Model and Existing Solutions}

\subsection{Massive MIMO Channel Model}
\begin{figure}
\center
\includegraphics[scale=0.45]{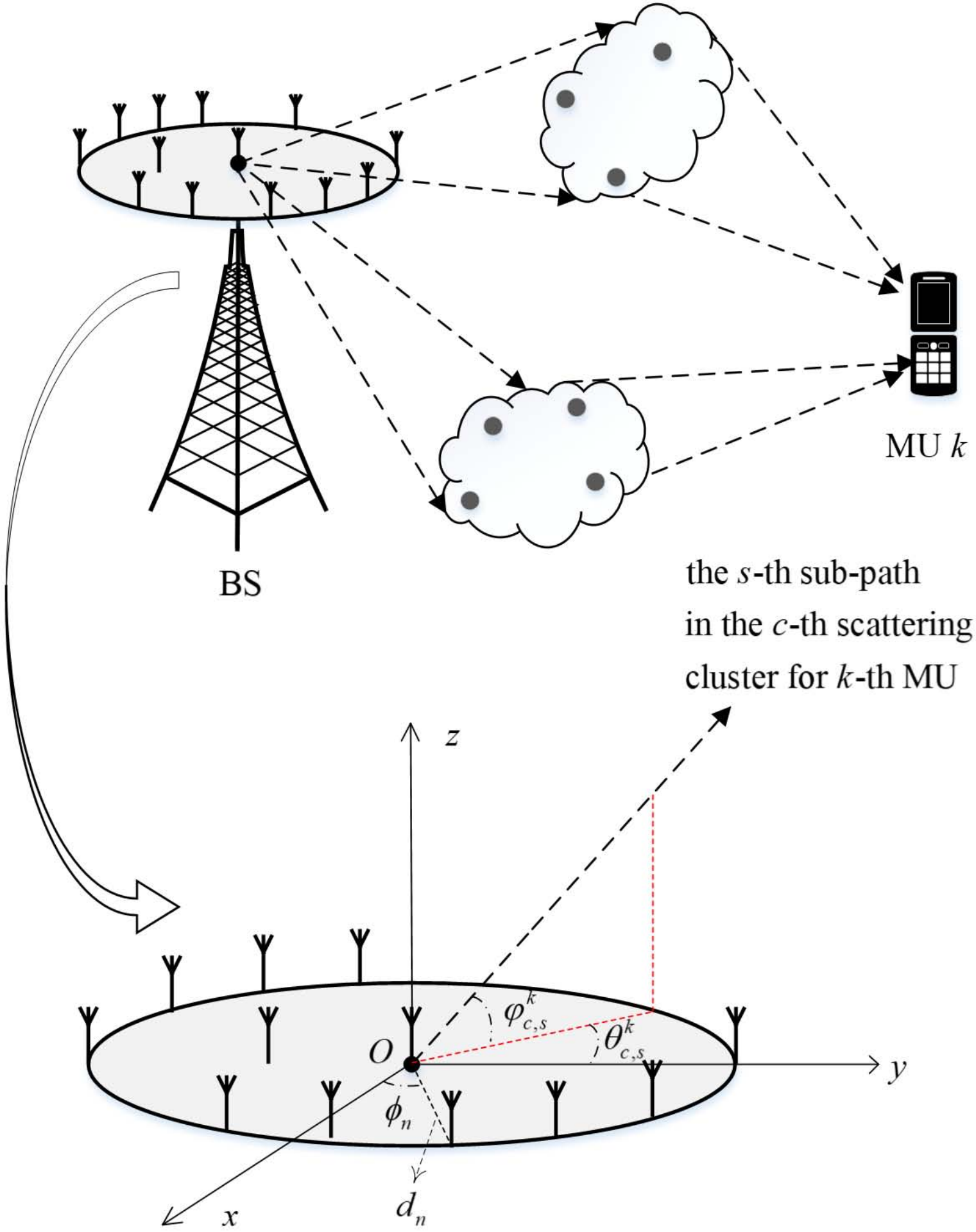}
\caption{Illustration of downlink channel model for a massive MIMO system. Note that we define the coordinate system with the X-Y plane being the plane of 2D array, regardless of how the plane of 2D array is placed at the BS (e.g., horizontally or
perpendicularly). }\label{figMIMO}
\end{figure}

Consider a massive MIMO system operating in FDD mode.  There is one BS with $N$ $(\gg1)$ antennas and $K$  MUs equipped with a single antenna. The array at the BS has an arbitrary geometry in the plane. Without loss of generality, we define the reference plane (the X-Y plane) to be the plane of 2D array, and set the origin of a polar coordinate system to be at the first element of the array,  as illustrated in Fig.~\ref{figMIMO}.
We consider a flat fading channel, and the downlink channel vector from the BS to the $k$-th user is given by \cite{tse2005fundamentals,3gpp}
\begin{align}
\h_k&= \sum_{c=1}^{N_c} \sum_{s=1}^{N_s} \xi_{c,s}^k \a(\theta_{c,s}^k, \varphi_{c,s}^k),\label{eqmo1}
\end{align}
where $N_c$ stands for the number of scattering clusters, $N_s$ stands for the number of sub-paths per scattering cluster, $\xi_{c,s}^k $  is the complex gain of the $s$-th sub-path in the $c$-th scattering cluster for the $k$-th MU,
$\theta_{c,s}^k$ and $\varphi_{c,s}^k$ are the corresponding  azimuth and elevation angles-of-departure (AoDs), respectively.
The steering vector $\a(\theta,\varphi)\in  \mathbb{C}^{N\times 1} $ is
\begin{align}
&\a(\theta,\varphi)
=[1,  e^{-j2\pi  \frac{d_2}{\lambda_d}\cos(\varphi)\sin(\theta- \phi_2) }, \notag\\
&~~~~~~~~~~~~~~~~~~~~~~~~~~~~\ldots, e^{-j2\pi  \frac{d_N }{\lambda_d}\cos(\varphi)\sin(\theta- \phi_N)} ]^T,
\label{eq2Dsteer}
\end{align}
where $(d_n, \phi_n)$ is the coordinates of the $n$-th sensor, and $\lambda_d$ is the wavelength of the downlink propagation.
For a linear array,  $\a(\theta,\varphi)$ can be simplified by
\begin{align}
\a(\theta) &=[1,  e^{-j2\pi \frac{d_2}{\lambda_d} \sin(\theta)}, \ldots, e^{-j2\pi\frac{d_N}{\lambda_d} \sin(\theta)}]^T.\label{eqabdod1}
\end{align}
Specifically, for a ULA, $\a(\theta)$ becomes
\begin{align}
\a(\theta) &=[1,  e^{-j2\pi \frac{d}{\lambda_d} \sin(\theta)}, \ldots, e^{-j2\pi\frac{(N-1)d}{\lambda_d} \sin(\theta)} ]^T,\label{eqdod1}
\end{align}
where $d$ stands for the distance between adjacent sensors.

According to the geometry-based stochastic channel model (GSCM) \cite{molisch2003geometry}, the number of scattering clusters $N_c$ is usually small, and the sub-paths associated with each scattering cluster are likely to concentrate in a small range around the line-of-sight (LOS) direction between the BS and the scattering cluster. Therefore, only a few dimensions in the angular domain are occupied, which, in return, brings a low dimensional representation for the massive MIMO channels \cite{chen2016pilot,gao2015spatially,shen2016compressed}.

\subsection{Review of Downlink Channel Estimation}

In this subsection, we review the state-of-the-art  DFT-based sparse channel estimation for the downlink channel in an FDD system.
Assume that the BS is equipped with a ULA, and it broadcasts a sequence of $T$ training pilot symbols, denoted by $\X\in  \mathbb{C}^{T\times N}$, for each MU to estimate the downlink channel. Then, the downlink received signal $\y_k\in \mathbb{C}^{T\times 1}$ at the $k$-th MU is given by
\begin{align}\label{downlinkrec}
\y_k= \X \h_k + \n_k,
\end{align}
where $\n_k\in \mathbb{C}^{T \times 1}$ stands for the additive complex Gaussian noise with each element being zero mean and variance $\sigma^2$ in the downlink, and $\tr (\X\X^H)=PTN$ with ${P}/{\sigma^2}$ measuring the training SNR.
Since the number of antennas $N$ at the BS is large, it is unlikely to obtain a robust recovery of $\h_k$ by using conventional channel estimation techniques, e.g., least squares (LS) method.
Recently, the emerging compressed sensing (CS) technique has given new interest in the problem of downlink channel estimation with limited training overhead. The main idea of these methods is to find a sparse representation of $\h_k$  in the DFT basis \cite{tse2005fundamentals}, i.e.,
\begin{align}
\h_k=\F\t_k \label{eq:dfth}
\end{align}
where $\F\in \mathbb{C}^{N\times N}$ denotes the DFT matrix   and $\t_k$ is the sparse representation channel vector.
Then, the received signal $\y_k$ in (\ref{downlinkrec})  can be formulated as
\begin{align}\label{eq-dly}
\y_k= \X \F\t_k + \n_k,
\end{align}
and the corresponding sparse signal recovery problem is given by
\begin{align}\label{dlnorm0}
\min_{\t_k} \|\t_k\|_0, ~~ \mathrm{subject~to}~   \|\y_k- \X \F\t_k\|_2\le \epsilon,
\end{align}
where $\epsilon$ is a constant determined by the upper bound of $\| \n_k\|_2$.
As $l_0$-norm is non-convex, it is usually relaxed by $l_1$-norm, i.e.,
\begin{align}\label{dlnorm1}
\min_{\t_k} \|\t_k\|_1, ~~ \mathrm{subject~to}~   \|\y_k- \X \F\t_k\|_2\le \epsilon.
\end{align}

\subsection{Challenges for the DFT-based Method}
In this subsection, we discuss challenges for the DFT-based method. Firstly, this method is applicable to ULAs only, which is explained as follows.
The DFT matrix can be written in the form of
\begin{align}
\F=\begin{bmatrix} \f\left(-\frac{1}{2}\right), &  \f\left(-\frac{1}{2} + \frac{1}{N} \right), & \ldots, & \f\left(\frac{1}{2} - \frac{1}{N} \right)
\end{bmatrix}
\end{align}
with
\begin{align}
\f(x)=\frac{1}{\sqrt{N}} \left[   1, e^{-j2\pi x}, \ldots,  e^{-j2\pi x (N-1)}\right]^T, \label{eqfeq}
\end{align}
which   provides a fixed  grid that uniformly covers  the range $[-\frac{1}{2}, \frac{1}{2}]$ with $N+1$ sampling points, i.e., $\left\{  -\frac{1}{2},~  (-\frac{1}{2}+  \frac{1}{N}),~ \ldots,   ~(\frac{1}{2} -  \frac{1}{N}) , \frac{1}{2}  \right\}$\footnote{Only the first $N$ points are used in the DFT matrix since $\f(-\frac{1}{2})=\f(\frac{1}{2})$.}.
As illustrated in (\ref{eqdod1}), the steering vectors of ULAs share the same structure with $\f(x)$.
For each sampling point (e.g., the $n$-th point), we can always  find a $\hat{\theta}_n$ in the angular domain such that
$
\frac{d}{\lambda_d} \sin(\hat{\theta}_n)= -\frac{1}{2} + \frac{n-1}{N}.
$
Hence, it is equivalent to saying the DFT basis actually provides a fixed sampling grid in the angular domain.
When the true azimuth AoDs $\theta_{c,s}^k$s
lie on (or, practically, close to) the sampling points  $\{ \hat{\theta}_1, \hat{\theta}_2,\ldots, \hat{\theta}_{N+1}  \}$, the channel vector  $\h_k$ definitely has a sparse representation in the DFT basis. Since the sparse property hinges strongly on the shared structure between the DFT basis and the ULA steering, the DFT-based method is applicable to ULAs only.

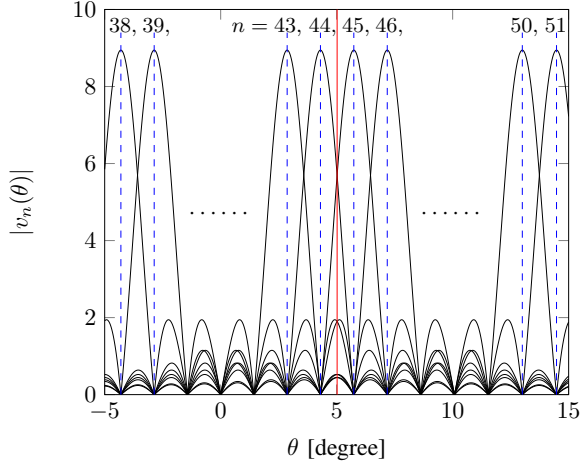
\begin{figure}
\center
\begin{tikzpicture}[scale=0.9]
\begin{axis}
[xlabel={$\theta$ [degree]},
ylabel={$|v_{n}(\theta)|$},
legend style={at={(0.88,0.93),
font=\footnotesize},
anchor=north,legend columns=1
}, xmin=-5,ymin=0,xmax=15,ymax=10]
\addplot[black] file {Fig_leak/38.txt} ;
\addplot[black] file {Fig_leak/39.txt} ;
\addplot[black] file {Fig_leak/43.txt} ;
\addplot[black] file {Fig_leak/44.txt} ;
\addplot[black] file {Fig_leak/45.txt} ;
\addplot[black] file {Fig_leak/46.txt} ;
\addplot[black] file {Fig_leak/50.txt} ;
\addplot[black] file {Fig_leak/51.txt} ;
\addplot[style= red] coordinates { (5.0198,0)   (5.0198,10) };
\addplot[style= dashed,blue] coordinates { (2.8660,0)     (2.8660,9.4) };  
\addplot[style= dashed,blue] coordinates { (4.3012,0)     (4.3012,9.4) };
\addplot[style= dashed,blue] coordinates { (5.7392,0)     (5.7392,9.4) };
\addplot[style= dashed,blue] coordinates { (7.1808,0)     (7.1808,9.4) };
\addplot[style= dashed,blue] coordinates { (-4.3012 ,0) (-4.3012 ,9.4) };
\addplot[style= dashed,blue] coordinates { (-2.8660 ,0) (-2.8660 ,9.4) };
\addplot[style= dashed,blue] coordinates { (13.0029 ,0) (13.0029 ,9.4) };
\addplot[style= dashed,blue] coordinates { (14.4775 ,0) (14.4775 ,9.4) };
\node (a) at (axis cs:-4.2012, 9.55) {\small $38,$ };
\node (a) at (axis cs:-2.7660, 9.55) {\small $39,$ };
\node (a) at (axis cs:2, 9.55) {\small $n=43,$ };
\node (a) at (axis cs:4.4012, 9.55) {\small $44,$ };
\node (a) at (axis cs:5.8392, 9.55) {\small $45,$ };
\node (a) at (axis cs:7.2808, 9.55) {\small $46,$ };
\node (a) at (axis cs:13.1029, 9.55) {\small $50,$ };
\node (a) at (axis cs:14.5775, 9.55) {\small $51,$ };
\node (b) at (axis cs:0,4.7) {$\cdots\cdots$ };
\node (b) at (axis cs:10,4.7) {$\cdots\cdots$ };
\end{axis}
\end{tikzpicture}
\caption{Illustration of problem of energy leakage for the DFT  basis, where the  sampling  points for the DFT basis in the angular domain are denoted by the dotted blue lines, the true azimuth AoD at $\theta= 5.0198^\circ$ is denoted by the red line, and the distance between the red line and the nearest dotted blue line is called the direction mismatch.
}\label{figleg}
\end{figure}

The other shortcoming of the DFT-based method is that it always has a performance loss, even for ULA systems, due to the leakage of energy.  As will be illustrated shortly, the leakage of energy caused by direction mismatch is unavoidable in practice.
According to  (\ref{eq:dfth}), we have
\begin{align}
\t_k=\F^H \h_k&=  \sum_{c=1}^{N_c}  \sum_{s=1}^{N_s} \xi_{c,s}^k \v(\theta_{c,s}^k),\label{eqdfw}
\end{align}
where $\v(\theta_{c,s}^k)= \F^H  \a(\theta_{c,s}^k)$. Then, the $n$-th element of $\v(\theta)$ can be calculated as 
\begin{align}
v_{n}(\theta)=& \frac{1}{\sqrt{N}} \sum_{i=0}^{N-1}  e^{j2\pi i (-\frac{1}{2}+\frac{ n-1 }{N})}          e^{-j2\pi \frac{i d }{\lambda_d} \sin(\theta)} \notag\\
=& \frac{1}{\sqrt{N}} \frac{1- e^{j2\pi(\frac{n-1}{N}-\frac{1}{2}- \frac{d}{\lambda_d}\sin(\theta))N} }
{1- e^{j2\pi(\frac{n-1}{N}-\frac{1}{2}- \frac{d}{\lambda_d}\sin(\theta)   )   }}\notag\\
=& \frac{1}{\sqrt{N}} \frac{\sin(\pi \varrho(\theta) N)}{ \sin(\pi \varrho(\theta))}  e^{j\pi \varrho(\theta) (N-1)}, \label{eqfte}
\end{align}
where $\varrho(\theta) = \frac{n-1}{N}-\frac{1}{2}- \frac{d}{\lambda_d}\sin(\theta) $.  Clearly, the modulus of  $v_{n}(\theta)$ (denoted as $|v_n(\theta)|$)
is a periodic function w.r.t. $\theta$, and it achieves the maximum value at $\theta=\hat\theta_n$.
If the true azimuth AoDs are located on the predefined  points $\{ \hat{\theta}_1, \hat{\theta}_2,\ldots, \hat{\theta}_{N+1}  \}$, there is no energy leakage.  In practice, however, direction mismatch is unavoidable because signals usually come from random directions.
Any direction mismatch will result in the leakage of energy in the DFT basis. Fig.~\ref{figleg} shows an example of the energy leakage,  where the ULA is of size $N=80$ and the inter-antenna spacing is a half wavelength. For an off-grid azimuth AoD (e.g, $\theta^*=5.0198^\circ$),  there is a very serious energy leakage, where both $|v_{44}(\theta^*)|$ and $|v_{45}(\theta^*)|$ have large values, as well as some significant values with  $|v_{43}(\theta^*)|$ and $|v_{46}(\theta^*)|$.

To achieve a better sparse representation, \cite{ding2015channel,ding2015compressed,ding2016dictionary}  applied an overcomplete DFT basis, which corresponds to a  denser sampling grid covering the  angular domain with more points.
However, if the grid is not sufficiently dense, the overcomplete DFT method may still lead to a high direction mismatch.
In order to solve the problem of direction mismatch, as well as apply the sparse channel estimation method to more general array geometry, we  propose an efficient SBL-based off-grid method for downlink channel estimation.
In the following, we first focus on a simple case, where the BS is equipped with a linear array, and thus only the azimuth angle is involved in the steering vector. This simple case  will help to simplify the algorithm design and link the proposed off-grid method with the state-of-the-art methods that are usually applicable to ULAs only.
After that, we extend the proposed off-grid method to an arbitrary 2D-array geometry, where both the azimuth and elevation angles are presented in the steering vector. Finally,  we exploit angular reciprocity to further improve channel estimation performance.

\section{Off-Grid Downlink Channel Estimation for Linear Array}
In this section, we will propose an efficient SBL-based off-grid method for downlink channel estimation with a linear array, which includes a ULA as a special case.
For ease of exposition, we proceed as follows.
We begin  by introducing a model-based off-grid basis to handle the direction mismatch  for a linear array. Then, we apply this off-grid model  to the  downlink channel estimation, and an in-exact MM algorithm is provided, as well as its convergence analysis.

\subsection{Off-Grid Basis for Massive MIMO Channels}

For ease of notation, we drop the MU's index $k$ and denote the true azimuth AoDs as $\{\theta_l,l=1,2,\ldots, L\}$, where $L=N_cN_s$. Let $\hat{\bm\vartheta}=\{\hat{\vartheta}_{l}\}_{l=1}^{\hat{L}}$ be a fixed sampling grid that  uniformly covers  the angular  domain $[-\frac{\pi}{2}, \frac{\pi}{2}]$, where $\hat L$ denotes the number of grid points.
If the grid is fine enough such that all the true DOAs  $\theta_l$s, $l=1,2,\ldots, L$, lie on (or practically close to) the grid, we can use the following model for $\h$:
\begin{align}\label{Hmodeln1}
\h= \A \w,
\end{align}
where $\A= \begin{bmatrix} \a ( \hat{\vartheta}_{1} ), & \a(\hat{\vartheta}_{2}), & \ldots, & \a (\hat{\vartheta}_{\hat L})
\end{bmatrix} \in \mathbb{C}^{N\times \hat L}  $, $\a(\theta)$ is a steering vector for a linear array [defined in (\ref{eqabdod1})],
and $\w \in \mathbb{C}^{\hat{L}\times 1} $ is a sparse vector whose non-zero elements correspond to the true directions at $\{\theta_l,l=1,2,\ldots, L\}$. For example, if the $\hat l$-th element of $\w$ is nonzero and the corresponding true direction is $\theta_l$, then we have $ \theta_l= \hat{\vartheta}_{\hat l}$.
Note that $\A$ includes the DFT basis as a special case.

As mentioned in Section II-B, the assumption of the true directions being located
on the predefined spatial grid is usually invalid in practice.  To handle the direction mismatch, we adopt an off-grid model. Specifically, if $\theta_l \notin \{\hat\vartheta_i\}_{i=1}^{\hat{L}}$ and $\hat\vartheta_{n_l}, n_l\in\{1,2,\ldots, \hat{L}\}$, is the nearest grid point to $\theta_l$,  we write $\theta_l$ as
\begin{align}
\theta_l= \hat\vartheta_{n_l} + \beta_{n_l},\label{eq-offg}
\end{align}
where $\beta_{n_l}$ corresponds to the off-grid gap. Using (\ref{eq-offg}), we have
$
\a\left(\theta_l \right)=
\a(\hat\vartheta_{n_l}+\beta_{n_l}).
$
Then, $\h$ can be rewritten as
\begin{align}\label{Hmodeln123}
\h= \A(\bm\beta) \w,
\end{align}
where $\bm\beta=[\beta_1, \beta_2,\ldots, \beta_{\hat L}]^T$, $\A(\bm\beta) = [ \a(\hat\vartheta_{1}+\beta_{1}),  \a(\hat\vartheta_{2}+\beta_{2}),\ldots, \a(\hat\vartheta_{\hat L}+\beta_{\hat L})  ] $, and
$$\beta_{n_l}=\begin{cases}\theta_l - \hat\vartheta_{n_l},  &l=1,2,\ldots, L   \\
 0,   & \mathrm{otherwise}\end{cases}.$$
Note that with the off-grid basis, the model can significantly  alleviate  the direction mismatch because there always exists some $\beta_{n_l}$ making (\ref{eq-offg}) hold exactly.
The received signal $\y$ in (\ref{downlinkrec}) can be rewritten by
\begin{align}\label{dlofmodel}
\y= \X \A(\bm\beta) \w + \n = \bm\Phi(\bm\beta)\w + \n,
\end{align}
where $\bm\Phi(\bm\beta)\triangleq\X \A(\bm\beta)$.
Since the coefficient vector $\bm\beta$ is unknown, the current $l_1$-norm minimization algorithm can not be applied to the off-grid channel model (\ref{dlofmodel}) directly. To jointly recover the sparse signal and refine the grid points, we adopt the
SBL algorithm \cite{tipping2001sparse,ji2008bayesian}, which is one
of the most popular approaches for sparse recovery and perturbation calibration.
Theoretical and empirical results show that  SBL methods can achieve enhanced
performance over $l_1$ regularized optimization (please also refer to our simulations). In the following, we will discuss how to jointly recover the sparse signal and refine the grid.

\subsection{Sparse Bayesian Learning Formulation}

Under the assumption of circular symmetric complex Gaussian noises, we have
\begin{align}\label{eq-yat}
p(\y | \w, \alpha, \bm\beta) =\mathcal{CN}(\y | \bm\Phi(\bm\beta) \w, \alpha^{-1}\I),
\end{align}
where $\alpha= \sigma^{-2}$ stands for the noise precision. Since $\alpha$ is usually unknown, we model it as a Gamma hyperprior
$p(\alpha)=\Gamma(\alpha;1+a,b)$, where we set $a,b\rightarrow 0$ as in \cite{tipping2001sparse,ji2008bayesian} so as to obtain a broad hyperprior.
We assume a noninformative uniform prior for $\bm\beta$:
\begin{align}
\bm\beta \thicksim U \left( \left[-\frac{\pi}{2}, \frac{\pi}{2}\right]^{\hat L}  \right).
\end{align}
Following the commonly used sparse Bayesian model \cite{tipping2001sparse}, we further assign a non-stationary Gaussian prior distribution with a distinct precision $\gamma_i$ for each element of $\w$.
Letting $\bm{\gamma} =[\gamma_1, \gamma_2,\ldots, \gamma_{\hat{L}}]^T$, we have
\begin{align}
p(\w|\bm{\gamma})=  \mathcal{CN}(\w | \bm{0}, \mathrm{diag}(\bm{\gamma}^{-1} ) ).\label{eq:mos}
\end{align}
Similarly, we model $\gamma_i$s as independent Gamma distributions, i.e.,
\begin{align}
p(\bm{\gamma})= \prod_{i=1}^{\hat{L}}  \Gamma(\gamma_i;1+a, b ).
\end{align}
This two-stage hierarchical prior gives  
\begin{align}
p(\w)=&\int_{0}^\infty   \mathcal{CN}(\w | \bm{0}, \mathrm{diag}(\bm{\gamma}^{-1} )  ) p(\bm{\gamma}) d\bm{\gamma}\notag\\
\propto & \prod_{i=1}^{\hat{L}}  \left( b+ |w_i|^2  \right)^{-(a+\frac{3}{2})},  \label{eqnewspp}
\end{align}
which is  recognized as encouraging sparsity due to the heavy tails and sharp peak at zero with a small $b$ \cite{tipping2001sparse,wipf2004sparse}.
In fact, it can be shown that finding a MAP estimate of $\w$ with the prior (\ref{eqnewspp}) is equivalent to  finding the minimum $l_0$-norm solution using FOCUSS with $p\rightarrow0$ \cite{rao1999affine}.
This explains why  SBL methods can achieve enhanced performance over the $l_1$-norm-based methods.
Since directly finding the aforementioned MAP estimate of $\w$ is difficult,  SBL methods introduce a two-stage hierarchical prior to get around the problematic MAP estimate. We refer  interested readers to Section V of \cite{wipf2004sparse} for details.

It is worth noting that the precisions $\gamma_l$s in (\ref{eq:mos}) fully indicate the support of $\w$. For example, if $\gamma_l$ is large, the $l$-th element of $\w$ tends to zero; otherwise, the value of the $l$-th  element is significant. As a consequence, once we obtain the precision vector $\bm\gamma$, as well as the off-grid gap $\bm\beta$, the estimated downlink channel $\h^e$ can be obtained by
\begin{align}
\h^e= \A_{\Omega}(\bm\beta) \left(\bm\Phi_{\Omega}(\bm\beta)\right)^{\dag} \y,
\end{align}
where $\Omega=\mathrm{supp}(\w)$.
Therefore, in the rest part of this section, we only need to focus on  finding the optimal $\bm\gamma$ and $\bm\beta$. As the noise precision $\alpha$ is still unknown, we find the  most-probable values $\alpha^\star$, $\bm\gamma^\star$ and $\bm\beta^\star $ together by maximizing the  posteriori $p(\alpha,\bm{\gamma},\bm\beta|\y)$, i.e.,
\begin{align}
(\alpha^\star, \bm\gamma^\star, \bm\beta^\star ) =&  \arg \max_{\alpha, \bm\gamma, \bm\beta} p(\alpha,\bm{\gamma},\bm\beta|\y), \end{align}
or, equivalently,
\begin{align}
(\alpha^\star, \bm\gamma^\star, \bm\beta^\star ) =&  \arg \max_{\alpha, \bm\gamma, \bm\beta}   \ln p(\y, \alpha,\bm{\gamma},\bm\beta). \label{eqEMp1}
\end{align}
The above objective is a high-dimensional non-convex function.
It is difficult to directly use the gradient ascent method on the original objective function because gradient ascent is known to have a slow convergence speed, and moreover, the gradient of the original objective function has no closed-form expression.
To overcome this challenge, we propose a novel in-exact block MM algorithm to find a stationary point of (\ref{eqEMp1}).

\subsection{Overview of the In-exact Block MM Algorithm}
The principle behind the block MM algorithm is to iteratively construct a  continuous surrogate function (lower bound) for the objective function
$\ln p(\y, \alpha,\bm{\gamma}, \bm\beta )$,  and then alternately maximize the surrogate function with  respect to $\alpha$, $\bm{\gamma}$ and $\bm\beta$.  The surrogate function is chosen such that the alternating maximization w.r.t. each variable has a closed-form/simple solution.

\begin{figure}
\center
\includegraphics[scale=0.28]{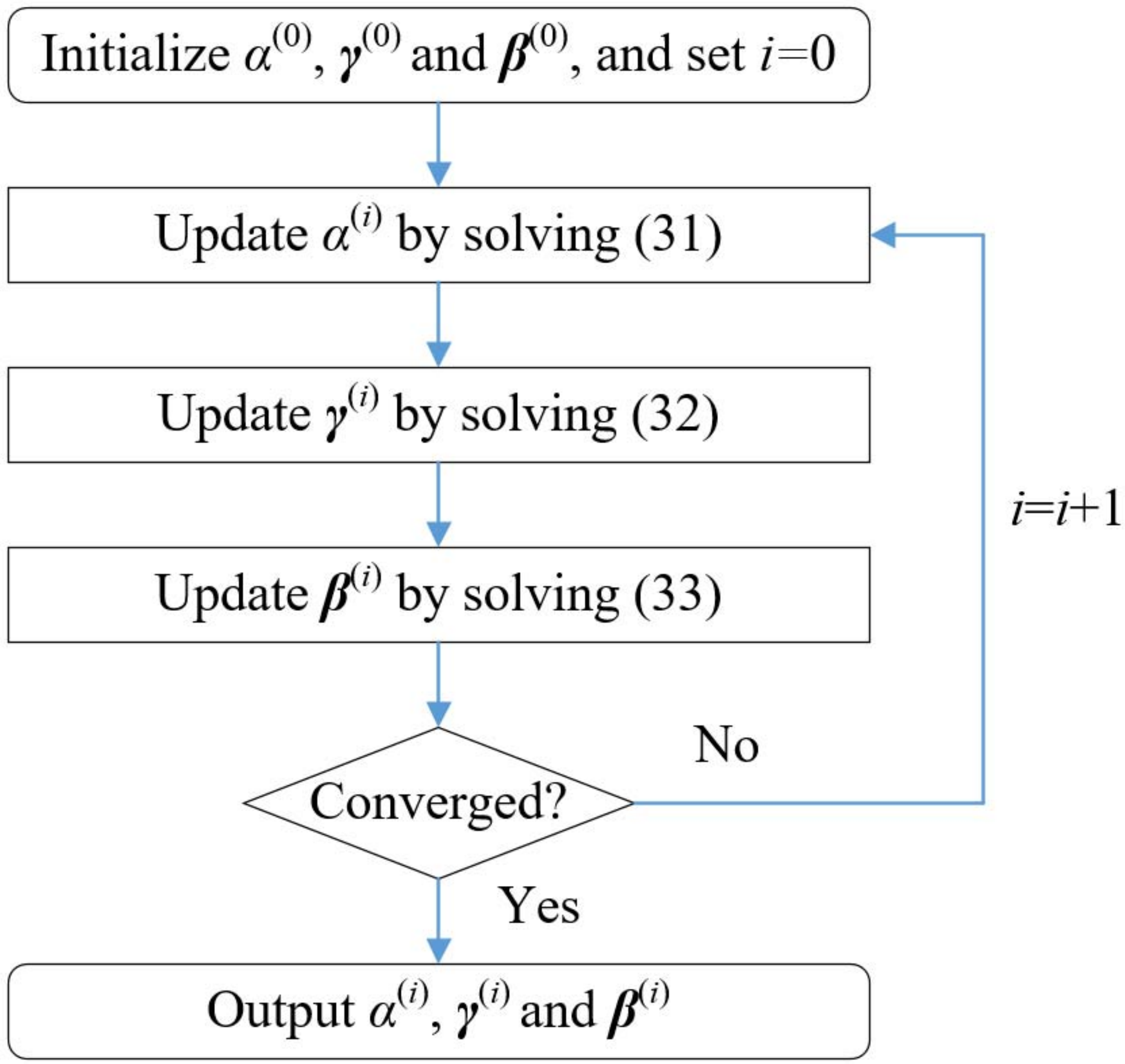}
\caption{The overall flow of the block MM algorithm.}\label{figFlow}
\end{figure}

Specifically, let  $\mathcal{U}(\alpha,\bm{\gamma}, \bm\beta| \dot{\alpha},\dot{\bm{\gamma}},\dot{\bm\beta})$ be the surrogate function constructed at some fixed point $(\dot{\alpha},\dot{\bm{\gamma}},\dot{\bm\beta})$ which satisfies  the following properties:
\begin{align}
&\mathcal{U}(\alpha,\bm{\gamma}, \bm\beta| \dot{\alpha},\dot{\bm{\gamma}},\dot{\bm\beta} ) \le \ln p(\y, \alpha,\bm{\gamma}, \bm\beta ), ~ \forall \alpha,\bm{\gamma}, \bm\beta, \label{eqMM1}\\
&\mathcal{U}(\dot{\alpha},\dot{\bm{\gamma}},\dot{\bm\beta} | \dot{\alpha},\dot{\bm{\gamma}},\dot{\bm\beta} ) = \ln p(\y, \dot{\alpha},\dot{\bm{\gamma}},\dot{\bm\beta} ),\label{eqMM2}\\
&\left.\frac{\partial \mathcal{U}(\alpha,\dot{\bm{\gamma}}, \dot{\bm\beta}| \dot{\alpha},\dot{\bm{\gamma}},\dot{\bm\beta})}{\partial \alpha}\right|_{\alpha=\dot{\alpha}} =  \left.\frac{\partial  \ln p(\y,  \alpha, \dot{\bm{\gamma}},\dot{\bm\beta} ) }{\partial \alpha}\right|_{\alpha=\dot{\alpha}},\label{eqMM3}\\
&\left.\frac{\partial \mathcal{U}(\dot{\alpha},{\bm{\gamma}}, \dot{\bm\beta}| \dot{\alpha},\dot{\bm{\gamma}},\dot{\bm\beta})}{\partial \bm\gamma}\right|_{\bm\gamma=\dot{\bm\gamma}} =  \left.\frac{\partial  \ln p(\y,  \dot{\alpha}, {\bm{\gamma}},\dot{\bm\beta} ) }{\partial \bm\gamma}\right|_{\bm\gamma=\dot{\bm\gamma}},\label{eqMM4}\\
&\left.\frac{\partial \mathcal{U}(\dot{\alpha},\dot{\bm{\gamma}}, {\bm\beta}| \dot{\alpha},\dot{\bm{\gamma}},\dot{\bm\beta})}{\partial \bm\beta}\right|_{\bm\beta=\dot{\bm\beta}} =  \left.\frac{\partial  \ln p(\y,  \dot{\alpha}, \dot{\bm{\gamma}},{\bm\beta} ) }{\partial\bm\beta}\right|_{\bm\beta=\dot{\bm\beta}}.\label{eqMM5}
\end{align}
Then, we update  $\alpha,\bm{\gamma}, and \bm\beta$ as
\begin{align}
\alpha^{(i+1)}&= \arg \max_{\alpha}   \mathcal{U}(\alpha,\bm{\gamma}^{(i)}, \bm\beta^{(i)}| \alpha^{(i)},\bm{\gamma}^{(i)},\bm\beta^{(i)} ),\label{eqM1}\\
\bm{\gamma}^{(i+1)}&= \arg \max_{\bm{\gamma}}   \mathcal{U}(\alpha^{(i+1)},\bm{\gamma}, \bm\beta^{(i)}| \alpha^{(i+1)},\bm{\gamma}^{(i)},\bm\beta^{(i)} ),\label{eqM2}\\
\bm\beta^{(i+1)}&= \arg \max_{\bm\beta}   \mathcal{U}(\alpha^{(i+1)},\bm{\gamma}^{(i+1)}, \bm\beta| \alpha^{(i+1)},\bm{\gamma}^{(i+1)},\bm\beta^{(i)} ),\label{eqM3}
\end{align}
where $(\cdot)^{(i)}$ stands for the $i$-th iteration.
The overall flow of the block MM algorithm is given in Fig.~\ref{figFlow}.
The update rules (\ref{eqM1})--(\ref{eqM3}) guarantee the  convergence of the block MM algorithm as follows.

\noindent\textbf{Lemma~1.} The update rules (\ref{eqM1})--(\ref{eqM3}) give a  non-decreasing sequence $\ln p(\y,  \alpha^{(i)},\bm{\gamma}^{(i)},\bm\beta^{(i)} )$, $i=1,2,3,\ldots$
\begin{proof}
See Appendix A.
\end{proof}

In the block MM algorithm, we need to obtain the optimal solutions for the maximization problems in (\ref{eqM1})--(\ref{eqM3}). However, the maximization problem w.r.t. $\bm\beta$ in (\ref{eqM3}) is non-convex and it is difficult to find its optimal solution. Therefore, in this paper, we use an in-exact MM algorithm where $\bm\beta^{(i+1)}$ is obtained by applying a simple one-step update.
In the following, we will discuss the choice of the surrogate function and the hyperparameter updates for $\alpha$, $\bm{\gamma}$ and $\bm\beta$, respectively.
Despite the in-exact update for $\bm\beta$, we will prove that the in-exact block MM algorithm still converges to a stationary solution of the optimization problem (\ref{eqEMp1}).

\subsection{Detailed Implementations}
To update $\alpha$, $\bm{\gamma}$ and $\bm\beta$, we first have to choose an appropriate surrogate function $\mathcal{U}(\alpha,\bm{\gamma}, \bm\beta| \cdot,\cdot,\cdot)$ that satisfies the properties mentioned in (\ref{eqMM1})--(\ref{eqMM5}).
Inspired by the expectation-maximization (EM) algorithm \cite{dempster1977maximum}, we use the corresponding lower
bound function as the surrogate function; i.e.,
for  any fixed point $(\dot{\alpha},\dot{\bm{\gamma}},\dot{\bm\beta})$,  we construct the surrogate function as
\begin{align}
&\mathcal{U}(\alpha,\bm{\gamma}, \bm\beta| \dot{\alpha},\dot{\bm{\gamma}},\dot{\bm\beta})\notag\\
=&\int p(\w| \y, \dot{\alpha},\dot{\bm{\gamma}},\dot{\bm\beta} ) \ln \frac{ p(\w, \y, \alpha,\bm{\gamma},\bm\beta) }{ p(\w| \y, \dot{\alpha},\dot{\bm{\gamma}},\dot{\bm\beta}  ) }  d \w, \label{eqsur}
\end{align}
and we have the following lemma.

\noindent\textbf{Lemma~2.} All the properties in  (\ref{eqMM1})--(\ref{eqMM5}) hold true with the  surrogate function $\mathcal{U}(\alpha,\bm{\gamma}, \bm\beta| \cdot,\cdot,\cdot )$ given in (\ref{eqsur}).
\begin{proof}
See Appendix B.
\end{proof}

Note that, from (\ref{eq-yat}) and (\ref{eq:mos}), $p(\w| \y, \alpha,\bm{\gamma},\bm\beta )$ is complex Gaussian \cite{tipping2001sparse,yang2013off}:
\begin{align}
p(\w|\y,\alpha,\bm{\gamma}, \bm\beta ) =\mathcal{CN}\left(\w| \bm{\mu}(\alpha,\bm{\gamma}, \bm\beta),  \bm{\Sigma}(\alpha,\bm{\gamma}, \bm\beta) \right),\label{eq-pdfs}
\end{align}
where
\begin{align*}
\bm\mu (\alpha,\bm{\gamma}, \bm\beta) &= \alpha \bm\Sigma(\alpha,\bm{\gamma}, \bm\beta) \bm\Phi^H(\bm\beta) \y , \\
\bm\Sigma(\alpha,\bm{\gamma}, \bm\beta) &=  \left( \alpha \bm\Phi^H(\bm\beta)\bm\Phi(\bm\beta)  + \mathrm{diag}(\bm{\gamma} )     \right)^{-1}.
\end{align*}

With $\mathcal{U}(\alpha,\bm{\gamma}, \bm\beta| \cdot,\cdot,\cdot)$,
we discuss the hyperparameter  updates for $\alpha$, $\bm{\gamma}$ and $\bm\beta$, respectively, as follows.

\subsubsection{\underline{{Update for $\alpha$}}} The maximization problem in (\ref{eqM1})  has a simple and closed-form solution:

\noindent\textbf{Lemma~3.}  The optimization problem (\ref{eqM1}) has a unique solution:
\begin{align}
\alpha^{(i+1)} = \frac{T + a}{ b + \eta(\alpha^{(i)},\bm{\gamma}^{(i)},\bm\beta^{(i)} ) }, \label{equpv1}
\end{align}
where
\begin{align*}
\eta(\alpha,\bm{\gamma},\bm\beta)
&=\tr\left( \bm\Phi(\bm\beta)\bm\Sigma(\alpha,\bm{\gamma},\bm\beta) \bm\Phi^H(\bm\beta) \right)\\
&~~~~~~~~~~~~~~~~~~+
\left\| \y  -  \bm\Phi(\bm\beta)\bm\mu(\alpha,\bm{\gamma},\bm\beta)  \right\|_2^2.
\end{align*}
\begin{proof}
See Appendix~C.
\end{proof}

\subsubsection{\underline{Update for $\bm{\gamma}$}}
The maximization problem in (\ref{eqM2}) also has a simple and closed-form solution:

\noindent\textbf{Lemma~4.}  The optimization problem (\ref{eqM2}) has a unique solution:
\begin{align}
\gamma_l^{(i+1)}= \frac{a+1}{b+  \left[\bm\Xi(\alpha^{(i+1)},\bm{\gamma}^{(i)},\bm\beta^{(i)})\right]_{ll} }, ~\forall l,\label{equpv2}
\end{align}
where
\begin{align*}
&\bm\Xi(\alpha,\bm{\gamma},\bm\beta)=\bm\Sigma(\alpha,\bm{\gamma},\bm\beta)+\bm\mu(\alpha,\bm{\gamma},\bm\beta) \bm\mu^H(\alpha,\bm{\gamma},\bm\beta).
\end{align*}
\begin{proof}
See Appendix~D.
\end{proof}

\subsubsection{\underline{Update for $\bm{\beta}$}} Since the maximization problem (\ref{eqM3}) is non-convex and it is difficult to find its optimal solution, we apply gradient update on the objective function of (\ref{eqM3}) and  obtain a simple one-step update for $\bm\beta$.
We name the procedure of updating $\bm\beta$ as grid refining, because it is related with the modeling error caused by the off-grid gap.
The derivative of the objective function in (\ref{eqM3}) w.r.t.  $\bm\beta$ can be calculated as
\begin{align}
\bm{\zeta}^{(i)}_\beta=[\zeta^{(i)}(\beta_1),\zeta^{(i)}(\beta_2),\ldots, \zeta^{(i)}(\beta_{\hat L})]^T, \label{eqderbeta}
\end{align}
with
\begin{align}
\zeta^{(i)}(\beta_l)=& 2 \mathrm{Re}\left(  (\a' (\hat\vartheta_{ l}+ \beta_{ l}) )^H \X^H\X (\a (\hat\vartheta_{ l}+ \beta_{ l})) \right) \cdot c_1^{(i)}\notag\\
&~~~~~~~~~~~~+ 2 \mathrm{Re}\left(  (\a' (\hat\vartheta_{ l}+ \beta_{ l}))^H \X^H \c_2^{(i)} \right), \label{eqderbeta1}
\end{align}
where $c_1^{(i)}=-\alpha^{(i+1)}(\chi_{ll}^{(i)}+ |\mu_{l}^{(i)}|^2) $, $\c_2^{(i)}=\alpha^{(i+1)} ((\mu_{l}^{(i)})^* \y_{-l}^{(i)}   -\X \sum_{j\ne l} \chi_{jl}^{(i)} \a (\hat\vartheta_{j}+ \beta_{ j}) )$,
$\y_{-l}^{(i)}= \y -   \X \cdot\sum_{j\ne l } (\mu_{j}^{(i)} \cdot\a (\hat\vartheta_{j}+ \beta_{j})) $, $\a' (\hat\vartheta_{j}+\beta_{ l})= d \a(\hat\vartheta_{j}+\beta_{ l}) /{d \beta_{ l}} $, $\mu_{l}^{(i)}$ and $\chi_{jl}^{(i)}$ denote the $l$-th element and the $(j,l)$-th element of $\bm\mu(\alpha^{(i+1)},\bm{\gamma}^{(i+1)},\bm\beta^{(i)})$ and $\bm\Sigma(\alpha^{(i+1)},\bm{\gamma}^{(i+1)},\bm\beta^{(i)})$, respectively. The detailed derivation for (\ref{eqderbeta1}) can be found in Appendix~E.
It is clear that the optimal solution for $\bm\beta$ is hard to obtain.
Fortunately,  due to the convergence property illustrated in (\ref{eqi2}), we just have to find a suboptimal solution  that increases the value of the objective function step by step. The most popular numerical method is to update the value of $\beta_l$ in  the derivative direction, i.e.,
\begin{align}\label{equpv3}
\bm\beta^{(i+1)}= \bm\beta^{(i)}  + \Delta_{\beta} \cdot \bm{\zeta}^{(i)}_\beta,
\end{align}
where $\Delta_{\beta}$ is the stepsize. Here, we can resort to backtracking line search \cite{wright1999numerical} to determine the maximum stepsize $\Delta_{\beta}$,  which ensures that the objective value can be strictly decreased before reaching the stationary point. The complexity of choosing the right stepsize mainly depends on calculating the cost function. If the number of calculations of  the cost function in every backtracking line search is $R_b$, the computational complexity is $\mathcal{O}(R_b T \hat{L}^2)$ per iteration for parameter tuning. Note that the complexity in calculating $\bm{\zeta}$ is $\mathcal{O}(T N\hat{L})$  per iteration. This suggests the computational requirement of updating $\bm\beta$ is $\mathcal{O}(R_b T \hat{L}^2)$   per iteration, because $\hat{L}$ is usually larger than $N$.
To reduce the computational complexity, we  alternatively use a fixed stepsize to update $\bm\beta$:
\begin{align}\label{equpv3-fix}
\bm\beta^{(i+1)}= \bm\beta^{(i)}   +       \frac{r_\theta}{100} \cdot  \mathrm{sign} (  \bm{\zeta}_\beta^{(i)} ),
\end{align}
where $r_\theta= \pi/\hat L$ stands for the grid interval, and $\mathrm{sign}(\cdot) $  stands for the signum function whose computational complexity is negligible. The motivation for choosing this fixed stepsize comes from the fact that a tiny difference between the obtained angles and the true angles does not affect the channel estimation performance much. The term $\frac{r_\theta}{100}$ guarantees that the final gap is smaller than $1\%$ of $r_\theta$, and the (approximate) true values  may be attained  within $100$ iterations in the worst case.

%
%

Finally, following are some  practical  implementation tips for the proposed method.
Empirical evidence shows that  the proposed method usually converges within 30 iterations, and it remains very robust to the choice of initialization. We can simply set the initialization as follows:  $a=b=0.0001$, $\alpha^{(0)}=1$, $\bm\gamma^{(0)}=\bm{1}$, and $\bm\beta^{(0)}=\bm{0}$.
Note that MATLAB codes  have been made available online at
\underline{https://sites.google.com/site/jsdaiustc/publication}.


\subsection{Convergence Analysis and Discussion}

From Lemma~1, the sequence $\ln p(\y,  \alpha^{(i)},\bm{\gamma}^{(i)},\bm\beta^{(i)} )$, $i=1,2,3,\ldots$, is non-decreasing and it converges to a limit because the evidence function has the upper bound of 1. In the following, we further prove that the sequence of iterates generated by the algorithm converges to a stationary point.

\noindent\textbf{Theorem~5.} For the surrogate function defined in (\ref{eqsur}), if variables are iteratively updated by (\ref{equpv1}), (\ref{equpv2}) and (\ref{equpv3}), the iterates generated by the in-exact block MM algorithm converge to a stationary solution of the optimization problem (\ref{eqEMp1}).
\begin{proof}
 See Appendix~F.
\end{proof}

Next, we address the difference between the proposed in-exact block MM algorithm  and  the EM algorithm.
The standard SBL method usually exploits the EM algorithm to perform the Bayesian inference.
The EM algorithm iteratively constructs the same  lower bound  as in (\ref{eqsur}),
and simultaneously updates $\alpha$, $\bm{\gamma}$ and  $\bm\beta$ by
\begin{align}
&(\alpha^{(i+1)},\bm{\gamma}^{(i+1)}, \bm\beta^{(i+1)})\notag\\
=& \arg\max_{\alpha,\bm{\gamma}, \bm\beta} \mathcal{U}(\alpha,\bm{\gamma}, \bm\beta|\y, \alpha^{(i)},{\bm{\gamma}}^{(i)},{\bm\beta}^{(i)}) .\label{emem1}
\end{align}
The EM algorithm can find a local optimal solution and its convergence  can be guaranteed, if the joint maximization problem in (\ref{emem1}) is solvable. Unfortunately,  in our problem, (\ref{emem1}) is non-convex and  is intractable in the presence of $\bm\beta$.  Hence, the EM algorithm cannot be directly applied to our problem. 

One commonly used method to address this challenge is to first obtain a convex approximation of the non-convex problem (\ref{emem1}) using the linear approximation off-grid model \cite{yang2013off,liu2017off}, and then update $\alpha$, $\bm{\gamma}$ and  $\bm\beta$  by solving the resulting convex approximation problem
in each iteration. Specifically, by replacing the steering vector $\bm\a(\theta_l) = \bm\a(\hat\vartheta_{n_l} + \beta_{n_l})$ with the linear approximation
\begin{align}
\bm\a(\theta_l)\approx \bm\a(\hat\vartheta_{n_l}) + \beta_{n_l} \cdot\a' ( \hat\vartheta_{n_l}),\label{eq-offdoa}
\end{align}
the surrogate function in (\ref{eqsur}) becomes convex and  thus can be maximized efficiently.
However, we do not adopt this linear approximation method, because if the grid is not sufficiently fine, (\ref{eq-offdoa}) may lead to a large modeling error, and the final channel estimation performance will be poor.

Finally, we discuss the computational complexity of the proposed method, whose
 main computational burden is given as follows.
\begin{itemize}
  \item The complexities in calculating $\bm\Sigma(\alpha,\bm{\gamma}, \bm\beta)$ and $\bm\mu (\alpha,\bm{\gamma}, \bm\beta)$  in each iteration are $\mathcal{O}(T\hat{L}^2)$ and $\mathcal{O}(\hat{L}^2)$, respectively.
  \item The complexities in updating $\alpha$ and  $\bm\gamma$ in each iteration are $\mathcal{O}(T\hat{L}^2)$ and $\mathcal{O}(\hat{L})$, respectively.
   \item The complexity in updating $\bm\beta$ is $\mathcal{O}(T N\hat{L})$  per iteration if the fixed stepsize update is used.
\end{itemize}
This suggests the total computational requirement of the proposed method is  $\mathcal{O}(T \hat{L}^2)$ per iteration, because $\hat L$ is usually larger than $N$.

\section{Extension to Arbitrary 2D-Array Geometry}

In this section, we extend the proposed off-grid method to an arbitrary 2D-array geometry, where the steering vector $\a(\theta,\varphi)$ [defined in (\ref{eq2Dsteer})] contains both azimuth  and elevation angles. Following the convention in Section III, we adopt a fixed sampling grid $\hat{\bm\vartheta}=\{\hat{\vartheta}_{l}\}_{l=1}^{\hat{L}}$ to uniformly cover the azimuth domain $[-\pi, \pi]$, and define the off-grid gap $\bm\beta$ similarly to (\ref{Hmodeln123}).
Then, the received signal $\y$ in (\ref{downlinkrec}) can be rewritten by
\begin{align}\label{2D-dlofmodel}
\y= \bm\Phi(\bm\beta,\hat{\bm\varphi})\w + \n,
\end{align}
where $\bm\Phi(\bm\beta,\hat{\bm\varphi}) = \X \A(\bm\beta,\hat{\bm\varphi}) $, $\A(\bm\beta,\hat{\bm\varphi})=  [\a(\hat{\vartheta}_{1} +  \beta_1,\hat\varphi_1) ,\a(\hat{\vartheta}_{2} +  \beta_2,\hat\varphi_2),\ldots, \a(\hat{\vartheta}_{\hat L} +  \beta_{\hat L},\hat\varphi_{\hat L})  ]$, and
$$\hat\varphi_{n_l}=\begin{cases}\varphi_l,  &l=1,2,\ldots, L   \\
 0,   & \mathrm{otherwise}\end{cases}.$$
Note that the definition of $n_l$ can be found in (\ref{eq-offg}). Compared with (\ref{dlofmodel}), the only difference is that
the measurement matrix $\bm\Phi(\bm\beta,\hat{\bm\varphi})$ contains an additional unknown variable (i.e., $\hat{\bm\varphi}$).
In the following, we will show how to blend $\hat{\bm\varphi}$ with the proposed off-grid method.

In the sparse Bayesian learning formulation for the new model (\ref{2D-dlofmodel}), almost all the results in Section III-B remain unchanged, except that (\ref{eq-yat}) is replaced by
\begin{align}\label{eq-yat-2D}
p(\y | \w, \alpha, \bm\beta, \hat{\bm\varphi}) =\mathcal{CN}(\y | \bm\Phi(\bm\beta, \hat{\bm\varphi}) \w, \alpha^{-1}\I),
\end{align}
and the optimization problem (\ref{eqEMp1}) is modified  by
\begin{align}
(\alpha^\star, \bm\gamma^\star, \bm\beta^\star, \hat{\bm\varphi}^\star ) =&  \arg \max_{\alpha, \bm\gamma, \bm\beta, \hat{\bm\varphi}}   \ln p(\y, \alpha,\bm{\gamma},\bm\beta, \hat{\bm\varphi}). \label{eqEMp1-2D}
\end{align}
For ease of notation, let $\Theta\triangleq\{\alpha,\bm{\gamma},\bm\beta, \hat{\bm\varphi} \}$.
At some fixed point $\dot{\Theta}=\{ \dot{\alpha},\dot{\bm{\gamma}},\dot{\bm\beta}, \dot{\hat{\bm\varphi}}  \}$,  we construct the surrogate function as
\begin{align}
\mathcal{U}({\Theta} | \dot{\Theta} )=\int p(\w| \y,   \dot{\Theta}  ) \ln \frac{ p(\w, \y,  {\Theta}) }{ p(\w| \y, \dot{\Theta} ) }  d \w. \label{eqsurr2-2D}
\end{align}
Then, in the maximization step of the $(i+1)$-th iteration, we update  $\alpha,\bm{\gamma}, \bm\beta$ and $\hat{\bm\varphi}$ as
\begin{align}
\alpha^{(i+1)}&= \arg \max_{\alpha}   \mathcal{U}(\alpha,\bm{\gamma}^{(i)}, \bm\beta^{(i)},\hat{\bm\varphi}^{(i)}| {\Theta}^{(i)} ),\label{eqM1-2D}\\
\bm{\gamma}^{(i+1)}&= \arg \max_{\bm{\gamma}}   \mathcal{U}(\alpha^{(i+1)},\bm{\gamma}, \bm\beta^{(i)},\hat{\bm\varphi}^{(i)} |  {\Theta}^{(i)}_1  ),\label{eqM2-2D}\\
\bm\beta^{(i+1)}&= \arg \max_{\bm\beta}   \mathcal{U}(\alpha^{(i+1)},\bm{\gamma}^{(i+1)}, \bm\beta,\hat{\bm\varphi}^{(i)} |  {\Theta}^{(i)}_2   ),\label{eqM3-2D}\\
\hat{\bm\varphi}^{(i+1)}&= \arg \max_{\hat{\bm\varphi}}   \mathcal{U}(\alpha^{(i+1)},\bm{\gamma}^{(i+1)}, \bm\beta^{(i+1)},\hat{\bm\varphi} |  {\Theta}^{(i)}_3   ),\label{eqM4-2D}
\end{align}
where $\Theta^{(i)}_j$ denotes the first $j$ elements of $\Theta^{(i)}$ coming from the $(i+1)$-th iteration.
Applying the results in Section III-D directly, we can obtain the solutions to (\ref{eqM1-2D})--(\ref{eqM3-2D}):
\begin{align}
\alpha^{(i+1)} =& \frac{T + a}{ b + \eta(\alpha^{(i)},\bm{\gamma}^{(i)},\bm\beta^{(i)},\hat{\bm\varphi}^{(i)} ) }, \label{equpv1-2D}\\
\gamma_l^{(i+1)}=& \frac{a+1}{b+  \left[\bm\Xi(\alpha^{(i+1)},\bm{\gamma}^{(i)},\bm\beta^{(i)},\hat{\bm\varphi}^{(i)})\right]_{ll} }, ~\forall l,\label{equpv2-2D}\\
\bm\beta^{(i+1)}=& \bm\beta^{(i)}   +       \frac{r_\theta}{100} \cdot  \mathrm{sign} (  \bm{\zeta}_\beta^{(i)} ),\label{equpv3-2D}
\end{align}
where
$\eta(\alpha,\bm{\gamma},\bm\beta,\hat{\bm\varphi})$, $\bm\Xi(\alpha,\bm{\gamma},\bm\beta,\hat{\bm\varphi})$, and $\bm{\zeta}_\beta$ follow the same definitions as in Section~III-D, except for the tiny modification of adding the new variable $\hat{\bm\varphi}$ after $\bm\beta$ for all the equalities.

What remains is to obtain the update for $\hat{\bm\varphi}$.
The last maximization problem (\ref{eqM4-2D}) is  similar to  (\ref{eqM3-2D}), where the objective function w.r.t $\hat{\bm\varphi}$ is also non-convex. Hence, we apply the same  one-step update for $\hat{\bm\varphi}$ as in (\ref{equpv3-2D}).
Following similar procedures to those in Appendix-E, we can obtain the derivative of the objective function w.r.t $\hat\varphi_l$ as
\begin{align}
\bm{\zeta}^{(i)}_\varphi=[\zeta^{(i)}(\hat\varphi_1),\zeta^{(i)}(\hat\varphi_2),\ldots, \zeta^{(i)}(\hat\varphi_{\hat L})]^T, \label{eqderbeta-2D}
\end{align}
with
\begin{align}
&\zeta^{(i)}(\hat\varphi_l)=\notag\\
& 2 \mathrm{Re}\left(  (\a'_{\varphi} (\hat\vartheta_{ l}+ \beta_{ l}^{(i+1)}, \hat\varphi_l) )^H \X^H\X (\a (\hat\vartheta_{ l}+ \beta_{ l}^{(i+1)},\hat\varphi_l)) \right) \cdot c_{\varphi 1}^{(i)}\notag\\
&+ 2 \mathrm{Re}\left(  (\a'_{\varphi} (\hat\vartheta_{ l}+ \beta_{ l}^{(i+1)}, \hat\varphi_l))^H \X^H \c_{\varphi 2}^{(i)} \right), \label{eqderbeta1-2D}
\end{align}
where $c_{\varphi 1}^{(i)}=-\alpha^{(i+1)}(\chi_{\varphi,ll}^{(i)}+ |\mu_{\varphi,l}^{(i)}|^2) $, $\c_{\varphi 2}^{(i)}=\alpha^{(i+1)} ((\mu_{\varphi,l}^{(i)})^* \y_{\varphi-l}^{(i)}   -\X \sum_{j\ne l} \chi_{\varphi,jl}^{(i)} \a (\hat\vartheta_{j}+ \beta_{ j}^{(i+1)}, \hat\varphi_j) )$,
$\y_{\varphi-l}^{(i)}= \y -   \X \cdot\sum_{j\ne l } (\mu_{\varphi,j}^{(i)} \cdot\a (\hat\vartheta_{j}+ \beta_{ j}^{(i+1)}, \hat\varphi_j)) $, $\a'_\varphi (\hat\vartheta_{j}+\beta_{ j}^{(i+1)}, \hat\varphi_j)= d \a(\hat\vartheta_{j}+\beta_{ j}^{(i+1)}, \hat\varphi_j) /{d \hat\varphi_{ j}} $, $\mu_{\varphi,l}^{(i)}$ and $\chi_{\varphi,jl}^{(i)}$ denote the $l$-th element and the $(j,l)$-th element of $\bm\mu(\alpha^{(i+1)},\bm{\gamma}^{(i+1)},\bm\beta^{(i+1)}, \hat{\bm\varphi}^{(i)}   )$ and $\bm\Sigma(\alpha^{(i+1)},\bm{\gamma}^{(i+1)},\bm\beta^{(i+1)}, \hat{\bm\varphi}^{(i)} )$, respectively.
With (\ref{eqderbeta-2D}), we are able to update $\hat{\bm\varphi}$ similarly to (\ref{equpv3}).

There are some important tips for practical implementations.
We note that the  elevation angle $\varphi$ ranges from $-\pi/2$ to $\pi/2$, but it is sufficient to assume that $\varphi$ ranges from $0$ to $\pi/2$, because the steering vector contains $\cos\varphi$ only. Hence, we initialize each $\hat\varphi_l$ uniformly from $[0, \pi/2]$. To reduce the computational complexity, we use a fixed stepsize to update $\hat{\bm\varphi}$ [similarly to (\ref{equpv3-fix})]:
\begin{align}\label{equpv3-fix-2D}
\hat{\bm\varphi}^{(i+1)}= \hat{\bm\varphi}^{(i)}   +  \frac{\pi}{36} \cdot \max \left\{ (\rho)^{i}, 0.001  \right\} \cdot \mathrm{sign} (  \bm{\zeta}_\beta^{(i)} ),
\end{align}
where $0<\rho<1$ is a constant.\footnote{The maximum movement is about $\frac{\pi}{36}\sum_{i=1}^{\infty} (\rho)^i $. In order to cover the whole angle domain $[0,\pi/2]$, $\rho$ should be chosen to be from $0.9474$ to $1$.}
Here, we use a different stepsize, $\frac{\pi}{36} \cdot \max \left\{ (\rho)^{i}, 0.001  \right\} $ instead of the $r_\theta/100$ in (\ref{equpv3-fix}), because there is no grid  covering the elevation angle domain. The motivation for choosing such a stepsize comes from the fact that 1) the term $\frac{\pi}{36}$ guarantees that the true elevation angles can be approximately approached within $20$ iterations; and 2) the term $  (\rho)^{i} $ keeps the stepsize decreasing, so as to attain a sufficiently small value [which is no less than $\frac{\pi}{36}\cdot 0.001$ due to the constant term $0.001$ in (\ref{equpv3-fix-2D})].


\section{Channel Estimation with Angular Reciprocity}
For the downlink channel estimation, the training period $T$ could become a bottleneck of the recovery performance, because the dimension of the measurement vector $\y$ in (\ref{dlofmodel}) is determined by the training period $T$, while $T$ is usually much less than $N$.  The performance of the downlink channel estimation can be improved if we collect more useful information.
Inspired by the angular reciprocity used in \cite{ding2016dictionary}, we present an off-grid uplink-AoA-aided channel estimation method in this section.
Here we only take the linear array as an example, but its extension to an arbitrary 2D-array antenna geometry is straightforward.
Note that angular reciprocity is quite different from the commonly
used channel reciprocity in TDD systems. In the first subsection, we will  explain angular reciprocity in detail.

\begin{figure}
\center
\includegraphics[scale=0.20]{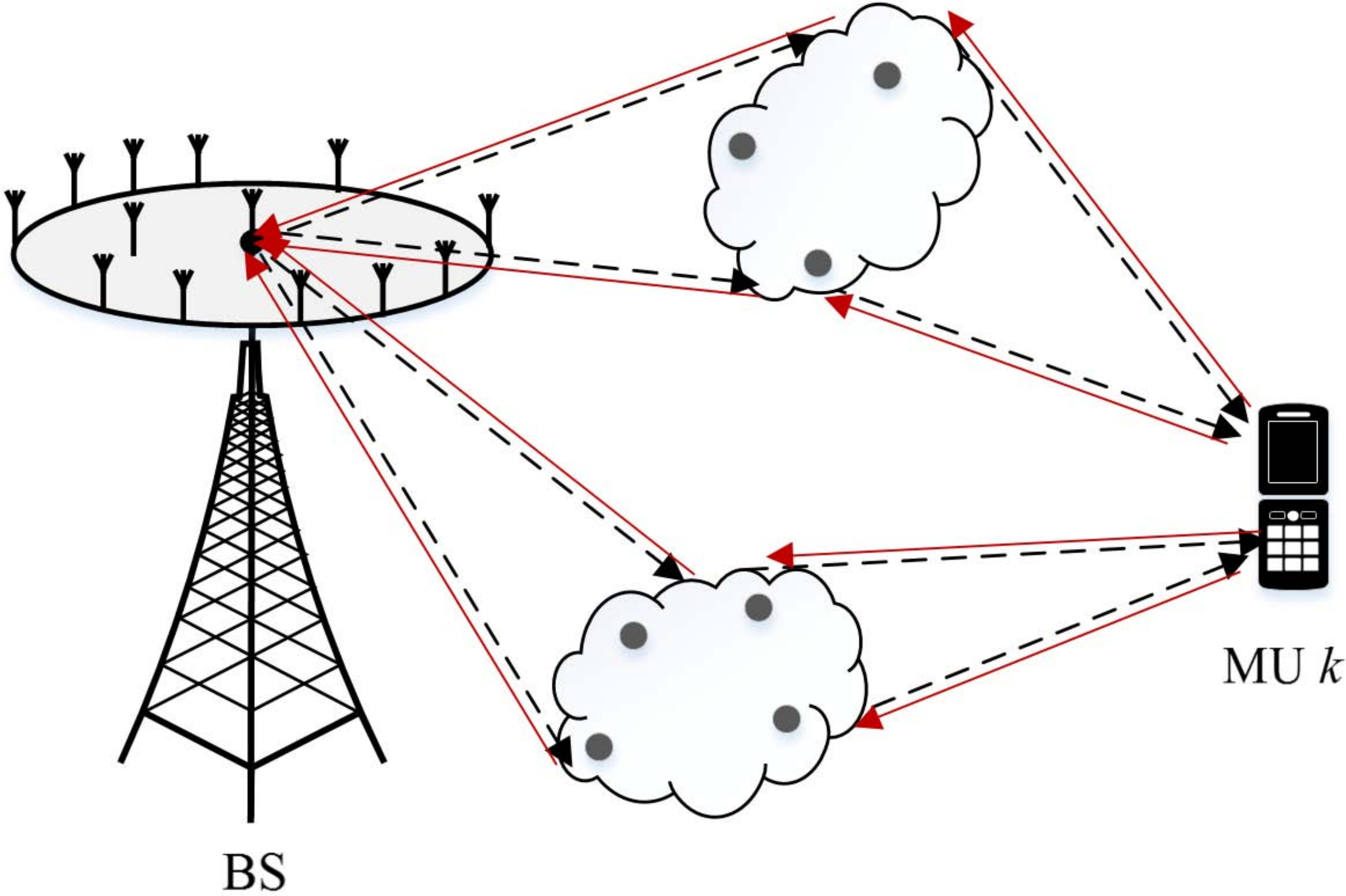}
\caption{Illustration of angular reciprocity for a massive MIMO system, where the downlink transmission and the uplink transmission are denoted by the dotted black lines and red lines, respectively.
}\label{figMIMO-uplink}
\end{figure}

\subsection{Angular Reciprocity}

Following the downlink channel model in Section~II-A, the uplink channel vector from the $k$-th user to the BS is given by
\begin{align}
\bar\h_k&= \sum_{c=1}^{N_c} \sum_{s=1}^{N_s} \bar\xi_{c,s}^k \bar\a(\bar\theta_{c,s}^k),
\end{align}
where $\bar\xi_{c,s}^k$ is similarly defined as $\xi_{c,s}^k$,  $\bar\theta_{c,s}^k$ is the corresponding azimuth angle-of-arrival (AoA), as illustrated in Fig.~\ref{figMIMO-uplink}, and $\bar\a(\theta)\in  \mathbb{C}^{N\times 1} $ is the steering vector for a linear array:
\begin{align}
\bar\a(\theta) &=[1,  e^{-j2\pi  \frac{d_2}{\lambda_u}\sin(\theta) }, \ldots, e^{-j2\pi  \frac{d_N }{\lambda_u}\sin(\theta)} ]^T,\notag
\end{align}
where  $\lambda_u$ is the wavelength of uplink propagation.
Usually, channel reciprocity does not hold in FDD systems because different frequency bands are used in  the downlink and uplink transmission.  However, if  the downlink and  uplink transmissions operate  closely in time, it is reasonable to have the following assumption:

\noindent\textbf{Assumption 6} (Angular Reciprocity \cite{ding2016dictionary}). The azimuth AoAs of signals for the $k$-th MU in the uplink transmission almost coincide with the azimuth AoDs of signals in the downlink transmission, i.e.,
\begin{align}
\theta_{c,s}^k= \bar\theta_{c,s}^k ,~ \forall k, c, s,
\end{align}
as illustrated in Fig.~\ref{figMIMO-uplink}.

To exploit the angular reciprocity,  Ding and Rao \cite{ding2016dictionary}  collected the downlink and uplink channel vectors for the $k$-th MU in pair
\begin{align}
\h_{k}=& {\F} {\t}_{k},\\
\bar\h_{k}=& {\F} {\bar\t}_{k},
\end{align}
where $\bar\t_k$ is the sparse representation of $\bar\h_k$ under the DFT basis for ULAs.
It is worth noting that  the steering vectors $\a(\theta_{c,s}^k)$ and $\bar\a(\bar\theta_{c,s}^k)$ are distinct if  different frequency bands are used. Hence, the  angular reciprocity between  the downlink and uplink transmissions does not bring a joint sparse structure for $\t_{k}$ and $\bar\t_{k}$. To get around this problem, they assume that the frequency duplex distance is not large (i.e., $\lambda_d \approx \lambda_u$). In this case, they approximately have
\begin{align}
\a(\theta_{c,s}^k)\approx\bar\a(\bar\theta_{c,s}^k),\label{adad4}
\end{align}
and then
\begin{align}
\mathrm{supp}(\t_{k})\approx\mathrm{supp}(\t_{k}). \label{eqsupp0}
\end{align}
As the joint sparse structure only holds approximately, it may results in  performance loss.
To handle this drawback, we will propose a joint off-grid model in the next subsection.

\subsection{Joint Off-Grid Model}

For the uplink channel estimation, assume that each MU broadcasts a sequence of $\bar T$ training pilot symbols, denoted by $\s_k \in \mathbb{C}^{\bar T\times 1}, k=1,2,\ldots, K$.  Then, the received signal $\bar\Y \in \mathbb{C}^{N\times \bar T}$ at the BS is given by
\begin{align}\label{upmode}
\bar\Y= \bar\H \S + \bar\N,
\end{align}
where $\bar\H=[\bar\h_{1}, \bar\h_{2}, \ldots, \bar\h_{K} ] \in \mathbb{C}^{N\times K}$ with $\bar\h_{k}$ being the channel vector for the $k$-th MU, $\S=[\s_1,\s_2,\ldots, \s_K]^T  \in \mathbb{C}^{K\times \bar T}$, and $\bar\N \in \mathbb{C}^{N\times \bar T}$ stands for the additive complex Gaussian noise with each element being zero mean and variance $\bar\sigma^2$ in the uplink. If the number of MUs is small, i.e., $K \le \bar T$, the uplink channel matrix $\H_u$ can be easily obtained by the conventional LS estimate, i.e.,
\begin{align}\label{Lsforup}
[\bar\h^{ls}_1, \bar\h^{ls}_2,\ldots, \bar\h^{ls}_K  ]  \triangleq\bar\Y  \S^{\dag}= \bar\H + \E,
\end{align}
or, equivalently,
\begin{align}
\bar\h^{ls}_k =& \bar\h_{k} + \e_k 
, ~~k=1,2,\ldots, K, \label{Lsfor1}
\end{align}
where $\bar\h^{ls}_k$ stands for the LS estimate of $\bar\h_k$ and $\E \triangleq[\e_1,\e_2,\ldots, \e_K]$ stands for the estimation error. If $\S$ consists of an orthogonal pilot sequence, $\E$ is i.i.d. Gaussian.
Compared with the requirement $T \ge N$ for the downlink channel estimation, it is much easier to meet the requirement $\bar T \ge K$ for the uplink channel estimation.

With (\ref{Lsfor1}) and (\ref{dlofmodel}), we are able to exploit  the sparse property of each MU independently.
We  drop the MU's index $k$ for ease of notation, and then the paired
sparse representation equalities can be rewritten as
\begin{align}
\y=&  \bm\Phi(\bm\beta)  \w + \n , \label{eqparyd} \\
\bar\h^{ls}=&  \bar{\bm\Phi}(\bm\beta)   \bar\w+ \e,
\end{align}
where (\ref{eqparyd}) coincides with (\ref{dlofmodel}),  $\bar{\bm\Phi}(\bm\beta)=  [ \bar\a(\hat\vartheta_{1}+\beta_{1}),  \bar\a(\hat\vartheta_{2}+\beta_{2}),\ldots, \bar\a(\hat\vartheta_{\hat L}+\beta_{\hat L})  ] $, and $\bar\w$ is the sparse representation of $\bar\h^{ls}$.
If the angular reciprocity holds,
it is easy to check that $\mathrm{supp}(\w)=\mathrm{supp}(\bar\w)$.
Different from the approximation method  \cite{ding2016dictionary} that hinges on the condition of (\ref{adad4}), our off-grid model guarantees  a jointly sparse structure from the angular domain directly, where neither approximation of $\lambda_d \approx \lambda_u$ nor the assumption of ULA at the BS is required.
In the following, we will show how to jointly recover the sparse vectors $\w$ and $\bar{\w}$ in the framework of SBL with the in-exact MM algorithm.
Since the  results  can be similarly derived by following the procedures in Section~III, detailed derivations are omitted for brevity.

\subsection{Sparse Bayesian Learning Formulation}

Under the assumption of circular symmetric complex Gaussian noises, we have
\begin{align}
p(\y | \w, \alpha, \bm\beta) &=\mathcal{CN}(\y | \bm\Phi(\bm\beta) \w, \alpha^{-1}\I),\\
p(\bar\h^{ls} | \bar{\w}, \bar{\alpha}, \bm\beta) &=\mathcal{CN}(\bar\h^{ls} | \bar{\bm\Phi}(\bm\beta) \bar\w, \bar{\alpha}^{-1}\I),
\end{align}
where $\bar{\alpha}$ stands for the noise precision of $\e$, which is further modeled as a Gamma hyperprior
$p(\bar{\alpha})=\Gamma(\bar{\alpha};~a,b)$. Recall that, in Section III-B, we have used $\bm\gamma$ to control the sparsity of $\w$ as follows:
\begin{align}\label{eq-an1}
p(\w|\bm{\gamma})=  \mathcal{CN}(\w | \bm{0},  \mathrm{diag}(\bm{\gamma^{-1}} ) ).
\end{align}
If we let $\bm{\tau}=[\tau_1, \tau_2,\ldots,\tau_{\hat L}]^T$ be a nonnegative vector  and
\begin{align}\label{eq-an2}
p(\bar\w|\bm{\gamma})=  \mathcal{CN}(\bar\w | \bm{0}, \mathrm{diag}((\bm{\gamma} \odot \bm{\tau})^{-1} ) ),
\end{align}
$\w$ and $\bar{\w}$ will share a joint sparse structure.
For example, $\gamma_l^{-1}$ tends to zero, so is $(\gamma_l \tau_l) ^{-1}$.
Therefore, (\ref{eq-an1}) and (\ref{eq-an2}) provide a mathematic representation of the angular reciprocity.
The estimated uplink channel $\bar\h^{ls}$ contains the AoA information in the uplink. The proposed method only exploits the azimuth AoA information in $\bar\h^{ls}$  to help in identifying the azimuth AoD in the downlink (downlink channel support) via the angular reciprocity in (\ref{eq-an1}) and (\ref{eq-an2}). The small scale fading information contained in the estimated uplink channel $\bar\h^{ls}$ cannot be exploited since the channel reciprocity does not hold for FDD systems.

\subsection{Bayesian Inference}

For ease of notation, let $\w^{a}\triangleq\{ \w,\bar{\w}\}$, $\y^{a}\triangleq\{\y, \bar\h^{ls}  \}$, $\Theta\triangleq\{\alpha, \bar{\alpha}, \bm\gamma,\bm\tau, \bm\beta \}$. We assume a noninformative uniform prior for $\bm{\tau}$.
At some fixed point $\dot{\Theta}=\{\dot{\alpha}, \dot{\bar\alpha}, \dot{\bm\gamma},\dot{\bm\tau}, \dot{\bm\beta} \}$,  we construct the surrogate function as
\begin{align}
\mathcal{U}({\Theta} | \dot{\Theta} )=\int p(\w^a| \y^a,   \dot{\Theta}  ) \ln \frac{ p(\w^a, \y^a,  {\Theta}) }{ p(\w^a| \y^a, \dot{\Theta} ) }  d \w^a, \label{eqsurr2}
\end{align}
where
\begin{align*}
&p(\w^a, \y^a,  {\Theta})= p(\y | \w, \alpha, \bm\beta)  p(\w|\bm{\gamma})
  p(\bar\h^{ls} | \bar{\w}, \bar{\alpha}, \bm\beta)\notag\\
 &~~~~~~~~~~~~~~~~~~~~~~~ \cdot p(\bar\w|\bm{\gamma},\bm\tau) p(\alpha)  p(\bar{\alpha})  p(\bm{\gamma})  p(\bm{\tau}) p(\bm\beta)
\end{align*}
and
\begin{align*}
&p(\w^{a}|\y^a, \Theta)=\mathcal{CN}(\w| \bm{\mu}(\alpha,\bm{\gamma}, \bm\beta) ,  \bm{\Sigma}(\alpha,\bm{\gamma}, \bm\beta) )\notag \\
&~~~~~~~~~~~~~~~~~~\cdot \mathcal{CN}(\bar\w|\bar{\bm\mu}(\bar{\alpha},\bm{\gamma}, \bm\tau, \bm\beta),  \bar{\bm{\Sigma}}(\bar{\alpha},\bm{\gamma}, \bm\tau, \bm\beta) ),
\end{align*}
with
\begin{align*}
&\bar{\mu} (\bar{\alpha},\bm{\gamma}, \bm\tau, \bm\beta) )= \bar{\alpha} \bar{\bm\Sigma}(\bar{\alpha},\bm{\gamma}, \bm\tau, \bm\beta) ) \bar{\bm\Phi}^H(\bm\beta) \bar\h^{ls},\\
&\bar{\bm\Sigma} (\bar{\alpha},\bm{\gamma}, \bm\tau, \bm\beta) )=  \left( \bar{\alpha} \bar{\bm\Phi}^H(\bm\beta)\bar{\bm\Phi}(\bm\beta)  + \mathrm{diag}(\bm{\gamma}\odot \bm\tau )     \right)^{-1}.
\end{align*}
Note that  $\bm\mu(\alpha,\bm{\gamma}, \bm\beta)$ and $\bm\Sigma(\alpha,\bm{\gamma}, \bm\beta)$ have been defined in (\ref{eq-pdfs}).

In the maximization step of the $(i+1)$-th iteration, we update  $\alpha, \bar{\alpha}, \bm\gamma,\bm\tau, \bm\beta $ as
\begin{align}
\alpha^{(i+1)}&= \arg \max_{\alpha}   \mathcal{U}(\alpha, \bar{\alpha}^{(i)}, \bm\gamma^{(i)},\bm\tau^{(i)}, \bm\beta^{(i)}| \Theta^{(i)} ),\label{eqU1}\\
\bar{\alpha}^{(i+1)}&= \arg \max_{\bar{\alpha}}   \mathcal{U}(\alpha^{(i+1)}, \bar{\alpha}, \bm\gamma^{(i)},\bm\tau^{(i)}, \bm\beta^{(i)}| \Theta^{(i)}_1 ),\label{eqU2}\\
\bm\gamma^{(i+1)}&= \arg \max_{\bm\gamma}   \mathcal{U}(\alpha^{(i+1)}, \bar{\alpha}^{(i+1)}, \bm\gamma,\bm\tau^{(i)}, \bm\beta^{(i)}| \Theta^{(i)}_2 ),\label{eqU3}\\
\bm\tau^{(i+1)}&= \arg \max_{\bm\tau}   \mathcal{U}(\alpha^{(i+1)}, \bar{\alpha}^{(i+1)}, \bm\gamma^{(i+1)},\bm\tau, \bm\beta^{(i)}| \Theta^{(i)}_3 ),\label{eqU4}\\
\bm\beta^{(i+1)}&= \arg \max_{\bm\beta}   \mathcal{U}(\alpha^{(i+1)}, \bar{\alpha}^{(i+1)}, \bm\gamma^{(i+1)},\bm\tau^{(i+1)}, \bm\beta| \Theta^{(i)}_4 ).\label{eqU5}
\end{align}
Extending Lemmas~3 and 4, the updates for $\alpha$, $\bar{\alpha}$, $\bm\gamma$ and $\bm\tau$ can be obtained as follows:
\begin{align}
&\alpha^{(i+1)} = \frac{T + a}{ b + \eta_d(\Theta^{(i)}) },\label{eq-newup1}\\
&\bar{\alpha}^{(i+1)} = \frac{N + a}{ b + \eta_u(\Theta^{(i)}_1) },\label{eq-newup2}\\
&\gamma_l^{(i+1)}=\frac{a+2}{b+ \left[\bm\Xi_d(\Theta^{(i)}_2) +  \tau_l\bm\Xi_u(\Theta^{(i)}_2)   \right]_{ll}  }, ~\forall l,\label{eq-newup3}\\
&\tau_l^{(i+1)}=\frac{1}{\left[\gamma_l ^{(i+1)} \bm\Xi_u(\Theta^{(i)}_3)   \right]_{ll}},~\forall l,\label{eq-newup4}
\end{align}
where
\begin{align*}
&\eta_d(\Theta)
=\tr\left( \bm\Phi(\bm\beta)\bm\Sigma(\alpha,\bm{\gamma},\bm\beta) \bm\Phi^H(\bm\beta) \right)\\
&~~~~~~~~~~~~~~~~~~~~~~~~~~~~~~~~+
\left\| \y  -  \bm\Phi(\bm\beta)\bm\mu(\alpha,\bm{\gamma},\bm\beta)  \right\|_2^2,\\
&\eta_u(\Theta)
=\tr\left( \bar{\bm\Phi}(\bm\beta)\bar{\bm\Sigma}(\bar{\alpha},\bm{\gamma},\bm\tau,\bm\beta) \bar{\bm\Phi}^H(\bm\beta) \right)\\
&~~~~~~~~~~~~~~~~~~~~~~~~~~~~+
\left\| \bar\h^{ls}  -  \bar{\bm\Phi}(\bm\beta)\bar{\mu}(\bar{\alpha},\bm{\gamma},\bm\tau,\bm\beta)  \right\|_2^2,\\
&\bm\Xi_d(\Theta)=\bm\Sigma(\alpha,\bm{\gamma},\bm\beta)+\bm\mu(\alpha,\bm{\gamma},\bm\beta) \bm\mu^H(\alpha,\bm{\gamma},\bm\beta),\\
&\bm\Xi_u(\Theta)=\bar{\bm\Sigma}(\bar{\alpha},\bm{\gamma},\bm\tau,\bm\beta)+\bar{\mu}(\bar{\alpha},\bm{\gamma},\bm\tau,\bm\beta) \bar{\mu}^H(\bar{\alpha},\bm{\gamma},\bm\tau,\bm\beta).
\end{align*}

%
%


Finally, we discuss how to refine the grid for  the uplink-AoA-aided channel estimation.
Ignoring the terms independent of $\bm\beta$, the objective function in (\ref{eqU5}) becomes
\begin{align}
&\mathcal{U}(\alpha^{(i+1)}, \bar{\alpha}^{(i+1)}, \bm\gamma^{(i+1)},\bm\tau^{(i+1)}, \bm\beta| \Theta^{(i)}_4 )\notag\\
=&  -\alpha^{(i+1)} \left\| \y  -  \bm\Phi(\bm\beta) \bm\mu^{(i)} \right\|_2^2\notag\\
&~~~~~~~~~~~~~~~~~~- \alpha^{(i+1)}  \tr\left(\bm\Phi(\bm\beta)\bm\Sigma^{(i)} \bm\Phi^H(\bm\beta)\right)\notag\\
&  -\bar{\alpha}^{(i+1)} \left\| \bar\h^{ls}  -  \bar{\bm\Phi}(\bm\beta) \bar{\bm\mu}^{(i)} \right\|_2^2\notag\\
&~~~~~~~~~~~~~~~~~~- \bar{\alpha}^{(i+1)}  \tr\left(\bar{\bm\Phi}(\bm\beta)\bar{\bm\Sigma}^{(i)}  \bar{\bm\Phi}^H(\bm\beta)\right),\label{equpthetanew}
\end{align}
where $\bm\mu^{(i)}\triangleq\bm\mu(\alpha^{(i+1)},\bm{\gamma}^{(i+1)},\bm\beta^{(i)})$, $\bm\Sigma^{(i)}\triangleq \bm\Sigma(\alpha^{(i+1)},\bm{\gamma}^{(i+1)},\bm\beta^{(i)})$, $\bar{\mu}^{(i)}\triangleq\bar{\mu}(\bar{\alpha}^{(i+1)},\bm{\gamma}^{(i+1)},\bm{\tau}^{(i+1)},\bm\beta^{(i)})$, and $\bar{\bm\Sigma}^{(i)}\triangleq\bar{\bm\Sigma}(\bar{\alpha}^{(i+1)},\bm{\gamma}^{(i+1)},\bm{\tau}^{(i+1)},\bm\beta^{(i)})$.
Calculating the derivative of (\ref{equpthetanew}) w.r.t.  $\beta_l$ leads to
\begin{align}
\zeta^{(i)}(\beta_l)=& 2 \mathrm{Re}\left(  (\a' (\hat\vartheta_{ l}+ \beta_{ l}))^H \X^H\X (\a (\hat\vartheta_{ l}+ \beta_{ l})) \right) \cdot c_{d1}^{(i)}\notag\\
&+ 2 \mathrm{Re}\left(  (\a' (\hat\vartheta_{ l}+ \beta_{ l}))^H \X^H \c_{d2}^{(i)} \right)\notag\\
&+2 \mathrm{Re}\left(  (\bar\a' (\hat\vartheta_{ l}+\beta_{ l}))^H (\bar\a (\hat\vartheta_{ l}+\beta_{ l})) \right) \cdot c_{u1}^{(i)}\notag\\
&+ 2 \mathrm{Re}\left(  (\bar\a' (\hat\vartheta_{ l}+\beta_{ l}))^H \c_{u2}^{(i)} \right),\label{eq-upadbeta}
\end{align}
where
$\bar\a' (\hat\vartheta_{j}+\beta_{ l})= d \bar\a(\hat\vartheta_{j}+ \beta_{ l}) /{d \beta_{ l}} $
$c_{d1}^{(i)}=-\alpha^{(i+1)}(\chi_{d,ll}^{(i)}+ |\mu_{d,l}^{(i)}|^2) $, $\c_{d2}^{(i)}=\alpha^{(i+1)} ((\mu_{d,l}^{(i)})^* \y_{d-l}^{(i)}   -\X \sum_{j\ne l} \chi_{d,jl}^{(i)} (\a (\hat\vartheta_{j}+\beta_{ j})) )$,
$c_{u1}^{(i)}=-\bar{\alpha}^{(i+1)}(\chi_{u,ll}^{(i)}+ |\mu_{u,l}^{(i)}|^2) $, $\c_{u2}^{(i)}=\bar{\alpha}^{(i+1)} ((\mu_{u,l}^{(i)})^* \h_{-l}^{(i)}   -\sum_{j\ne l} \chi_{u,jl}^{(i)} (\bar\a (\hat\vartheta_{j}+ \beta_{ j})) )$,
$\y_{d-l}^{(i)}= \y -   \X \cdot\sum_{j\ne l } (\mu_{d,j}^{(i)} \cdot\a (\hat\vartheta_{j}+\beta_{j}) ) $,
$\h_{-l}^{(i)}= \bar\h^{ls} -  \sum_{j\ne l } (\mu_{u,j}^{(i)} \cdot\bar\a (\hat\vartheta_{j}+\beta_{j}) ) $,
 $\mu_{d,l}^{(i)}$ ($\mu_{u,l}^{(i)}$) and $\chi_{d,jl}^{(i)}$ ($\chi_{u,jl}^{(i)}$) denote the $l$-th element and the $(j,l)$-th element of $\bm\mu^{(i)}$ ($\bar{\mu}^{(i)}$) and $\bm\Sigma^{(i)}$ ($\bar{\bm\Sigma}^{(i)}$), respectively.
With (\ref{eq-upadbeta}), we are able to update  $\bm\beta$  similarly as in (\ref{equpv3}) or (\ref{equpv3-fix}.)
Noth that following the same initializations mentioned in Section III-D, we set $\alpha^{(0)}=\bar\alpha^{(0)}=1$, $\bm\gamma^{(0)}=\bm\tau^{(0)}=\bm{1}$ and $\bm\beta^{(0)}=\bm{0}$ for the off-grid uplink-AoA-aided method.

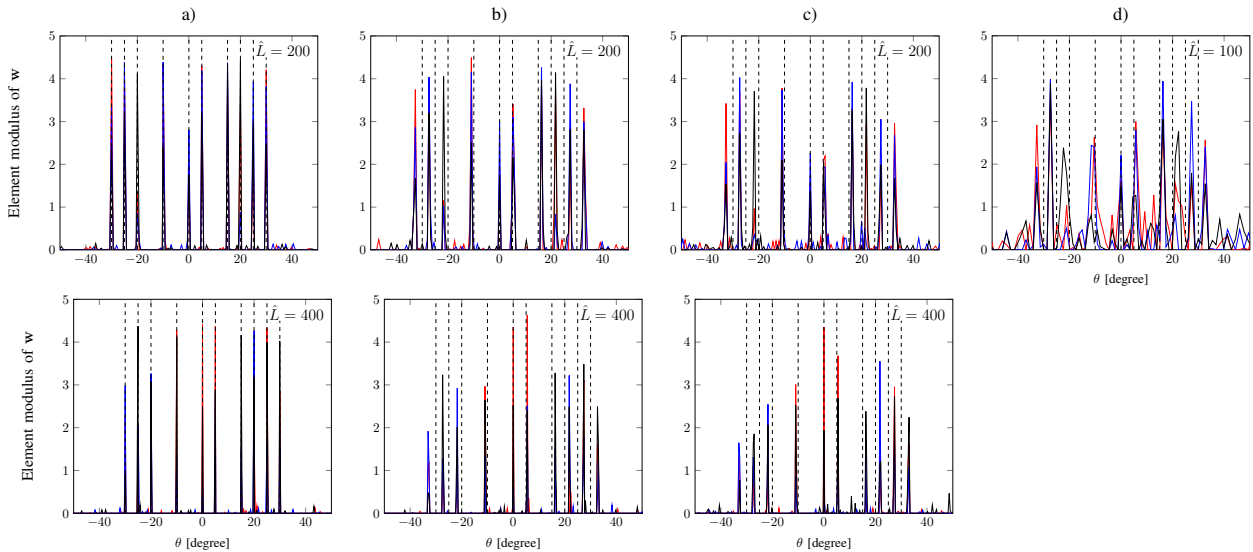
\begin{figure*}
\center
\begin{tikzpicture}[scale=0.5]
\begin{axis}
[xlabel={\textcolor[rgb]{1.00,1.00,1.00}{DOA [degree]}},
ylabel={\large Element modulus of $\w$},
legend style={at={(0.28,0.83),
font=\footnotesize},
anchor=north,legend columns=1
}, xmin=-50,ymin=0,xmax=50,ymax=5,title={\Large a)}]
\addplot[color=red] file {F1/R1/offgrid.txt} ;
\addplot[color=blue] file {F1/R2/offgrid.txt} ;
\addplot[color=black] file {F1/R3/offgrid.txt} ;
\addplot[style= dashed] coordinates { (-30,0) (-30,5) };
\addplot[style= dashed] coordinates { (-25,0) (-25,5) };
\addplot[style= dashed] coordinates { (-20,0) (-20,5) };
\addplot[style= dashed] coordinates { (-10,0) (-10,5) };
\addplot[style= dashed] coordinates {   (0,0)   (0,5) };
\addplot[style= dashed] coordinates {   (5,0)   (5,5) };
\addplot[style= dashed] coordinates {  (15,0)  (15,5) };
\addplot[style= dashed] coordinates {  (20,0)  (20,5) };
\addplot[style= dashed] coordinates {  (25,0)  (25,5) };
\addplot[style= dashed] coordinates {  (30,0)  (30,4.5) };
\node (a) at (axis cs:36.5,4.7) {\large $\hat L = 200$ };
\end{axis}
\end{tikzpicture}
~~
\begin{tikzpicture}[scale=0.5]
\begin{axis}
[xlabel={\textcolor[rgb]{1.00,1.00,1.00}{DOA [degree]}},
legend style={at={(0.28,0.83),
font=\footnotesize},
anchor=north,legend columns=1
}, xmin=-50,ymin=0,xmax=50,ymax=5,title={\Large b)}]
\addplot[color=red] file {F1/R1/SBL.txt} ;
\addplot[color=blue] file {F1/R2/SBL.txt} ;
\addplot[color=black] file {F1/R3/SBL.txt} ;
\addplot[style= dashed] coordinates { (-30,0) (-30,5) };
\addplot[style= dashed] coordinates { (-25,0) (-25,5) };
\addplot[style= dashed] coordinates { (-20,0) (-20,5) };
\addplot[style= dashed] coordinates { (-10,0) (-10,5) };
\addplot[style= dashed] coordinates {   (0,0)   (0,5) };
\addplot[style= dashed] coordinates {   (5,0)   (5,5) };
\addplot[style= dashed] coordinates {  (15,0)  (15,5) };
\addplot[style= dashed] coordinates {  (20,0)  (20,5) };
\addplot[style= dashed] coordinates {  (25,0)  (25,5) };
\addplot[style= dashed] coordinates {  (30,0)  (30,4.5) };
\node (a) at (axis cs:36.5,4.7) {\large $\hat L = 200$ };
\end{axis}
\end{tikzpicture}
~~
\begin{tikzpicture}[scale=0.5]
\begin{axis}
[xlabel={\textcolor[rgb]{1.00,1.00,1.00}{DOA [degree]}},
legend style={at={(0.28,0.83),
font=\footnotesize},
anchor=north,legend columns=1
}, xmin=-50,ymin=0,xmax=50,ymax=5,title={\Large c)}]
\addplot[color=red] file {F1/R1/OFFT.txt} ;
\addplot[color=blue] file {F1/R2/OFFT.txt} ;
\addplot[color=black] file {F1/R3/OFFT.txt} ;
\addplot[style= dashed] coordinates { (-30,0) (-30,5) };
\addplot[style= dashed] coordinates { (-25,0) (-25,5) };
\addplot[style= dashed] coordinates { (-20,0) (-20,5) };
\addplot[style= dashed] coordinates { (-10,0) (-10,5) };
\addplot[style= dashed] coordinates {   (0,0)   (0,5) };
\addplot[style= dashed] coordinates {   (5,0)   (5,5) };
\addplot[style= dashed] coordinates {  (15,0)  (15,5) };
\addplot[style= dashed] coordinates {  (20,0)  (20,5) };
\addplot[style= dashed] coordinates {  (25,0)  (25,5) };
\addplot[style= dashed] coordinates {  (30,0)  (30,4.5) };
\node (a) at (axis cs:36.5,4.7) {\large $\hat L = 200$ };
\end{axis}
\end{tikzpicture}
~~
\begin{tikzpicture}[scale=0.5]
\begin{axis}
[xlabel={$\theta$ [degree]},
legend style={at={(0.28,0.83),
font=\footnotesize},
anchor=north,legend columns=1
}, xmin=-50,ymin=0,xmax=50,ymax=5,title={\Large d)}]
\addplot[color=red] file {F1/R1/FFT.txt} ;
\addplot[color=blue] file {F1/R2/FFT.txt} ;
\addplot[color=black] file {F1/R3/FFT.txt} ;
\addplot[style= dashed] coordinates { (-30,0) (-30,5) };
\addplot[style= dashed] coordinates { (-25,0) (-25,5) };
\addplot[style= dashed] coordinates { (-20,0) (-20,5) };
\addplot[style= dashed] coordinates { (-10,0) (-10,5) };
\addplot[style= dashed] coordinates {   (0,0)   (0,5) };
\addplot[style= dashed] coordinates {   (5,0)   (5,5) };
\addplot[style= dashed] coordinates {  (15,0)  (15,5) };
\addplot[style= dashed] coordinates {  (20,0)  (20,5) };
\addplot[style= dashed] coordinates {  (25,0)  (25,5) };
\addplot[style= dashed] coordinates {  (30,0)  (30,5) };
\node (a) at (axis cs:36.5,4.7) {\large $\hat L = 100$ };
\end{axis}
\end{tikzpicture}
\begin{tikzpicture}[scale=0.5]
\begin{axis}
[xlabel={$\theta$ [degree]},
ylabel={\large Element modulus of $\w$},
legend style={at={(0.28,0.83),
font=\footnotesize},
anchor=north,legend columns=1
}, xmin=-50,ymin=0,xmax=50,ymax=5]
\addplot[color=red] file {F2/R1/offgrid.txt} ;
\addplot[color=blue] file {F2/R2/offgrid.txt} ;
\addplot[color=black] file {F2/R3/offgrid.txt} ;
\addplot[style= dashed] coordinates { (-30,0) (-30,5) };
\addplot[style= dashed] coordinates { (-25,0) (-25,5) };
\addplot[style= dashed] coordinates { (-20,0) (-20,5) };
\addplot[style= dashed] coordinates { (-10,0) (-10,5) };
\addplot[style= dashed] coordinates {   (0,0)   (0,5) };
\addplot[style= dashed] coordinates {   (5,0)   (5,5) };
\addplot[style= dashed] coordinates {  (15,0)  (15,5) };
\addplot[style= dashed] coordinates {  (20,0)  (20,5) };
\addplot[style= dashed] coordinates {  (25,0)  (25,5) };
\addplot[style= dashed] coordinates {  (30,0)  (30,4.5) };
\node (a) at (axis cs:36.5,4.7) {\large $\hat L = 400$ };
\end{axis}
\end{tikzpicture}
~~
\begin{tikzpicture}[scale=0.5]
\begin{axis}
[xlabel={$\theta$ [degree]},
legend style={at={(0.28,0.83),
font=\footnotesize},
anchor=north,legend columns=1
}, xmin=-50,ymin=0,xmax=50,ymax=5]
\addplot[color=red] file {F2/R1/SBL.txt} ;
\addplot[color=blue] file {F2/R2/SBL.txt} ;
\addplot[color=black] file {F2/R3/SBL.txt} ;
\addplot[style= dashed] coordinates { (-30,0) (-30,5) };
\addplot[style= dashed] coordinates { (-25,0) (-25,5) };
\addplot[style= dashed] coordinates { (-20,0) (-20,5) };
\addplot[style= dashed] coordinates { (-10,0) (-10,5) };
\addplot[style= dashed] coordinates {   (0,0)   (0,5) };
\addplot[style= dashed] coordinates {   (5,0)   (5,5) };
\addplot[style= dashed] coordinates {  (15,0)  (15,5) };
\addplot[style= dashed] coordinates {  (20,0)  (20,5) };
\addplot[style= dashed] coordinates {  (25,0)  (25,5) };
\addplot[style= dashed] coordinates {  (30,0)  (30,4.5) };
\node (a) at (axis cs:36.5,4.7) {\large $\hat L = 400$ };
\end{axis}
\end{tikzpicture}
~~
\begin{tikzpicture}[scale=0.5]
\begin{axis}
[xlabel={$\theta$ [degree]},
legend style={at={(0.28,0.83),
font=\footnotesize},
anchor=north,legend columns=1
}, xmin=-50,ymin=0,xmax=50,ymax=5]
\addplot[color=red] file {F2/R1/OFFT.txt} ;
\addplot[color=blue] file {F2/R2/OFFT.txt} ;
\addplot[color=black] file {F2/R3/OFFT.txt} ;
\addplot[style= dashed] coordinates { (-30,0) (-30,5) };
\addplot[style= dashed] coordinates { (-25,0) (-25,5) };
\addplot[style= dashed] coordinates { (-20,0) (-20,5) };
\addplot[style= dashed] coordinates { (-10,0) (-10,5) };
\addplot[style= dashed] coordinates {   (0,0)   (0,5) };
\addplot[style= dashed] coordinates {   (5,0)   (5,5) };
\addplot[style= dashed] coordinates {  (15,0)  (15,5) };
\addplot[style= dashed] coordinates {  (20,0)  (20,5) };
\addplot[style= dashed] coordinates {  (25,0)  (25,5) };
\addplot[style= dashed] coordinates {  (30,0)  (30,4.5) };
\node (a) at (axis cs:36.5,4.7) {\large $\hat L = 400$ };
\end{axis}
\end{tikzpicture}
\hspace{3.645cm}
\caption{Element modulus of $\w$ for three independent trials with $N=100$, $T=40$ and SNR$=10$ dB. The true azimuth AoDs are denoted by dotted lines.  a) Off-Grid; b) SBL; c) Overcomplete DFT; d) DFT.}\label{figss3}
\end{figure*}

\section{Simulation Results}

In this section, we conduct simulations to investigate the performance of our proposed methods.
The proposed methods
are compared with the following methods:
\begin{itemize}
\item \textbf{Baseline~1} (SBL): $\h_k$ is recovered using the standard SBL method \cite{tipping2001sparse} with the  dictionary $\A$ defined in (\ref{Hmodeln1}).
\item \textbf{Baseline~2} (DFT)): $\h_k$ is recovered using the  $l_1$-norm minimization algorithm \cite{Donoho2008,donoho2006compressed,candes2006robust} with a DFT basis.
\item \textbf{Baseline~3} (Overcomplete DFT)): $\h_k$ is recovered using the  $l_1$-norm minimization algorithm \cite{Donoho2008,donoho2006compressed,candes2006robust} with the dictionary $\A$ defined in (\ref{Hmodeln1}).
\item \textbf{Baseline~4} (Dictionary Learning)): $\h_k$ is recovered using the  method proposed in \cite{ding2016dictionary} with the  dictionary $\A$ defined in (\ref{Hmodeln1}).
\end{itemize}
Since the state-of-the-art DFT methods work for ULAs only, we first focus on simulations for ULAs, where
we use the 3GPP spatial channel model (SCM) \cite{3gpp} to generate the channel coefficients for an urban microcell.
The uplink frequency is $1980$~MHz, the downlink frequency is $2170$~MHz, and the inter-antenna spacing is $d=c/(2f_0)$, with $c$ being the light speed and $f_0=2000$~MHz.  Then, we run  simulations with the 3GPP 3D
channel model \cite{3gpp-3D-Model}, which provides a 2D array model involving both azimuth and
elevation angles. All the parameters of the 3D channel model follow 3D-UMa-NOLS (see Table 7.3-6 in \cite{3gpp-3D-Model}) and the downlink frequency is $2170$~MHz.
The normalized mean square error (NMSE) is defined as
\begin{align}
\frac{1}{M_c}\sum_{m=1}^{M_c} \frac{ \| {\h}_m^e - \h_m  \|_2^2   }{\| \h_m  \|_2^2},
\end{align}
where ${\h}_m^e$ is the estimate of $ \h_m $ at the $n$-th Monte Carlo trial and $M_c=200$ is the number of Monte Carlo trials.


\subsection{Recovered Channel Sparsity in the Angular Domain for ULA}

In Fig.~\ref{figss3}, we illustrate the effect of direction mismatch on the channel sparse representation performance for  different
channel estimation strategies. Consider a simple scenario where  a ULA with $100$ antennas at the BS is used to send the training pilot symbols with ten azimuth AoDs in total, which are simply denoted as $\theta_1=-30^\circ$, $\theta_2=-25^\circ$, $\theta_3=-20^\circ$, $\theta_4=-10^\circ$, $\theta_5=0^\circ$,
$\theta_6=5^\circ$, $\theta_7=15^\circ$, $\theta_8=20^\circ$, $\theta_9=25^\circ$, and $\theta_{10}=30^\circ$.
The training pilots are randomly generated with $T=40$ and the SNR is set to $10$ dB.
Fig.~\ref{figss3} shows the element modulus of the recovered  channel sparse representation $\w$, where
the number of grid points $\hat L$ is fixed to $200$ or $400$ for all the methods, except for the classical DFT method. It is observed that 1) the solution of the classical DFT method is not exactly sparse, and it has a significant performance loss
due to the leakage of energy over many bins; 2) the standard SBL method and the overcomplete DFT method can achieve better sparse representations, especially for a dense grid ($\hat L=400$), but direction mismatch always exists; and
3) our proposed off-grid method can greatly improve the sparsity and accuracy of the channel representation, and  the direction mismatch can be almost eliminated.

\subsection{Channel Estimation Performance Versus $T$ for ULA} 

In Fig.~\ref{fig-vsT1}, Monte Carlo trials are carried out to investigate the impact of the number of pilot symbols on the downlink channel estimation performance for ULA. Assume that the ULA at the BS is equipped with $150$ antennas, and the system supports ten MUs, where each MU has a single antenna. All the results are obtained by averaging over $200$ Monte Carlo channel realizations. Every  channel realization consists of $N_c=3$ random scattering clusters ranging from $-40^\circ$ to $40^\circ$, and each cluster contains $N_s=10$ sub-paths concentrated in a $20^\circ$ angular spread. The training pilots are randomly generated, the SNR is chosen as  $0$ dB or $10$ dB, and the number of grid points is fixed to $200$ for all but the DFT method. Fig.~\ref{fig-vsT1} shows the NMSE performance of the downlink channel estimate achieved by the different channel estimation strategies versus the number of training pilot symbols $T$. It can be seen that 1) the NMSEs of all the methods decrease as the number of training pilot symbols increases, and the DFT method gives the worst performance;
2) compared with the  DFT method, the state-of-the-art methods (the overcomplete DFT and dictionary learning method) can improve the NMSE performance, but the improvement is not significant (they are all worse than the standard SBL method); and 3) our proposed off-grid method always outperforms  the state-of-the-art methods, and the uplink-AoA-aided
method can further improve the performance of the downlink
channel estimation, as it collects more useful information than the off-grid method.

\begin{figure}
\center
\begin{tikzpicture}[scale=0.9]
\begin{semilogyaxis}[
 width=5cm,
ylabel={NMSE},grid=both, title={a)},
legend style={at={(0.77,1.55),font=\footnotesize},
anchor=north,legend columns=1}, xmin=30,xmax=100, ymin=0.015,x=3]
\addplot[mark=asterisk,blue]  coordinates{
  (    30 ,      0.8497)
  (    40 ,      0.5005)
  (    50 ,      0.2271)
  (    60 ,      0.0783)
  (    70 ,      0.0370)
  (    80 ,      0.0294)
  (    90 ,      0.0282)
  (   100 ,      0.0280)
};
\addplot[mark=o,red]  coordinates{
  (    30 ,     0.8177)
  (    40 ,     0.2839)
  (    50 ,     0.0867)
  (    60 ,     0.0414)
  (    70 ,     0.0297)
  (    80 ,     0.0266)
  (    90 ,     0.0262)
  (   100 ,     0.0246)
};
\addplot[mark=diamond]  coordinates{
  (    30 ,      0.8907)
  (    40 ,      0.5843)
  (    50 ,      0.3499)
  (    60 ,      0.1796)
  (    70 ,      0.0931)
  (    80 ,      0.0599)
  (    90 ,      0.0485)
  (   100 ,      0.0418)
};
\addplot[mark=pentagon] coordinates{
  (    30 ,       1.9449)
  (    40 ,       1.8026)
  (    50 ,       1.7678)
  (    60 ,       1.4384)
  (    70 ,       1.1977)
  (    80 ,       0.5126)
  (    90 ,       0.2200)
  (   100 ,       0.1078)
};
\addplot[mark=triangle] coordinates{
  (    30 ,       1.8245)
  (    40 ,       1.6507)
  (    50 ,       1.4876)
  (    60 ,       0.9087)
  (    70 ,       0.5345)
  (    80 ,       0.1351)
  (    90 ,       0.0596)
  (   100 ,       0.0407)
};
\addplot[mark=square] coordinates{
  (    30 ,      1.8056)
  (    40 ,      1.5411)
  (    50 ,      1.3982)
  (    60 ,      0.7627)
  (    70 ,      0.2765)
  (    80 ,      0.0899)
  (    90 ,      0.0509)
  (   100 ,      0.0391)
};
\legend{Off-Grid, Uplink-Aided, SBL , DFT, Overcomplete DFT, Dictionary Learning}
\end{semilogyaxis}
\end{tikzpicture}
~~
\begin{tikzpicture}[scale=0.9]
\begin{semilogyaxis}[xlabel={Number of training pilot symbols}, width=5cm,
ylabel={NMSE},grid=both, title={b)},
legend style={at={(0.77,1.55),font=\footnotesize},
anchor=north,legend columns=1}, xmin=30,xmax=100,x=3]
\addplot[mark=asterisk,blue]  coordinates{
  (    30 ,      0.8054)
  (    40 ,      0.4174)
  (    50 ,      0.1618)
  (    60 ,      0.0250)
  (    70 ,      0.0027)
  (    80 ,      0.0018)
  (    90 ,      0.0014)
  (   100 ,      0.0013)
};
\addplot[mark=o,red]  coordinates{
  (    30 ,      0.5937)
  (    40 ,      0.1159)
  (    50 ,      0.0141)
  (    60 ,      0.0033)
  (    70 ,      0.0018)
  (    80 ,      0.0014)
  (    90 ,      0.0012)
  (   100 ,      0.0010)
};
\addplot[mark=diamond]  coordinates{
  (    30 ,      0.8634)
  (    40 ,      0.5205)
  (    50 ,      0.3035)
  (    60 ,      0.1179)
  (    70 ,      0.0489)
  (    80 ,      0.0192)
  (    90 ,      0.0105)
  (   100 ,      0.0054)
};
\addplot[mark=pentagon] coordinates{
  (    30 ,      1.9352)
  (    40 ,      1.8001)
  (    50 ,      1.7068)
  (    60 ,      1.4230)
  (    70 ,      0.6973)
  (    80 ,      0.3322)
  (    90 ,      0.1536)
  (   100 ,      0.0808)
};
\addplot[mark=triangle] coordinates{
  (    30 ,       1.6978)
  (    40 ,       1.6577)
  (    50 ,       1.3311)
  (    60 ,       0.6882)
  (    70 ,       0.4619)
  (    80 ,       0.0506)
  (    90 ,       0.0253)
  (   100 ,       0.0115)
};
\addplot[mark=square] coordinates{
  (    30 ,       1.6917)
  (    40 ,       1.5435)
  (    50 ,       1.2153)
  (    60 ,       0.5697)
  (    70 ,       0.1439)
  (    80 ,       0.0438)
  (    90 ,       0.0188)
  (   100 ,       0.0094)
};
\end{semilogyaxis}
\end{tikzpicture}
\caption{NMSE of dowlink channel estimate versus the number of  training pilot symbols for ULA. a) SNR $=$ $0$ dB;  b) SNR $=$ $10$ dB.
 }\label{fig-vsT1}
\end{figure}
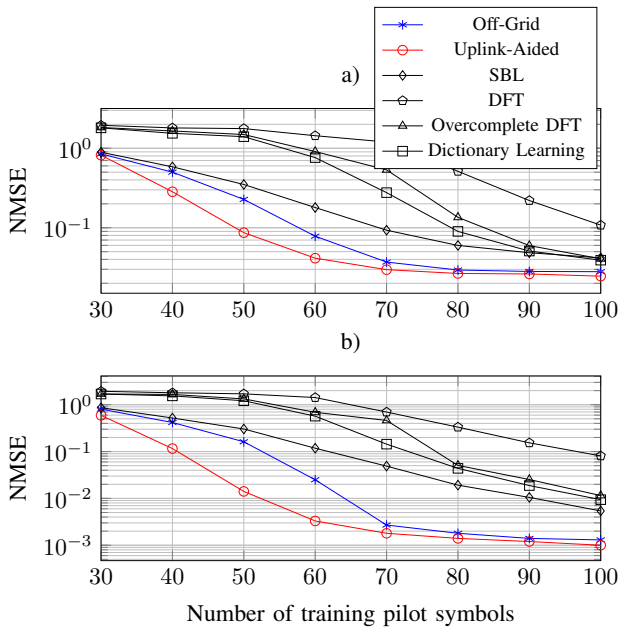


\begin{figure}
\center
\begin{tikzpicture}[scale=0.9]
\begin{semilogyaxis}[
width=5cm,
ylabel={NMSE},grid=both, title={a)},
legend style={at={(0.77,1.55),font=\footnotesize},
anchor=north,legend columns=1},xmin=150,xmax=400,ymin=0.015,x=0.839 ]
\addplot[mark=asterisk,blue]  coordinates{
  (   150,      0.0453)
  (   200,      0.0320)
  (   250,      0.0278)
  (   300,      0.0272)
  (   350,      0.0268)
  (   400,      0.0262)
};
\addplot[mark=o,red]  coordinates{
   (   150,      0.0327)
   (   200,      0.0260)
   (   250,      0.0234)
   (   300,      0.0236)
   (   350,      0.0229)
   (   400,      0.0220)
};
\addplot[mark=diamond]  coordinates{
  (   150,      0.1722)
  (   200,      0.0697)
  (   250,      0.0459)
  (   300,      0.0437)
  (   350,      0.0390)
  (   400,      0.0382)
};
\addplot[mark=pentagon] coordinates{
  (   150,      1.3893)
  (   200,      1.4421)
  (   250,      1.4160)
  (   300,      1.2616)
  (   350,      1.3315)
  (   400,      1.2904)
};
\addplot[mark=triangle] coordinates{
  (   150,     1.3893)
  (   200,     0.7993)
  (   250,     0.4984)
  (   300,     0.3897)
  (   350,     0.3009)
  (   400,     0.3099)
};
\addplot[mark=square] coordinates{
   (   150,     1.3893)
  (   200,      0.7993)
  (   250,      0.4984)
  (   300,      0.3897)
  (   350,      0.3009)
  (   400,      0.3099)
};
\legend{Off-Grid, Uplink-Aided, SBL , DFT, Overcomplete DFT, Dictionary Learning, Genie-aided LS}
\end{semilogyaxis}
\end{tikzpicture}
~~
\begin{tikzpicture}[scale=0.9]
\begin{semilogyaxis}[xlabel={Number  of  grid points},
width=5cm,
ylabel={NMSE},grid=both, title={b)},
legend style={at={(0.769,1.45),font=\footnotesize},
anchor=north,legend columns=1},xmin=150,xmax=400,ymax=5,ymin=0.0006,x=0.839 ]
\addplot[mark=asterisk,blue]  coordinates{
  (   150,     0.2767)
  (   200,     0.0163)
  (   250,     0.0037)
  (   300,     0.0024)
  (   350,     0.0015)
  (   400,     0.0014)
};
\addplot[mark=o,red]  coordinates{
   (   150,      0.0436)
   (   200,      0.0022)
   (   250,      0.0013)
   (   300,      0.0011)
   (   350,      0.0010)
   (   400,      0.0010)
};
\addplot[mark=diamond]  coordinates{
  (   150,      1.1409)
  (   200,      0.2262)
  (   250,      0.0771)
  (   300,      0.0399)
  (   350,      0.0277)
  (   400,      0.0171)
};
\addplot[mark=pentagon] coordinates{
  (   150,       1.6884)
  (   200,       1.7026)
  (   250,       1.6433)
  (   300,       1.5856)
  (   350,       1.6597)
  (   400,       1.6123)
};
\addplot[mark=triangle] coordinates{
  (   150,       3.9462)
  (   200,       1.7026)
  (   250,       1.1446)
  (   300,       0.7283)
  (   350,       0.5966)
  (   400,       0.5044)
};
\addplot[mark=square] coordinates{
   (   150,       3.9462)
  (   200,        1.7026)
  (   250,        1.1446)
  (   300,        0.7283)
  (   350,        0.5966)
  (   400,        0.5044)
};
\end{semilogyaxis}
\end{tikzpicture}
\caption{NMSE of downlink channel estimate versus the number of  grid points for ULA. a) $N=150$, $N_c=2$ and SNR $= 0$ dB;  b) $N=200$, $N_c=3$ and SNR= $10$ dB.
 }\label{figNg1}
\end{figure}
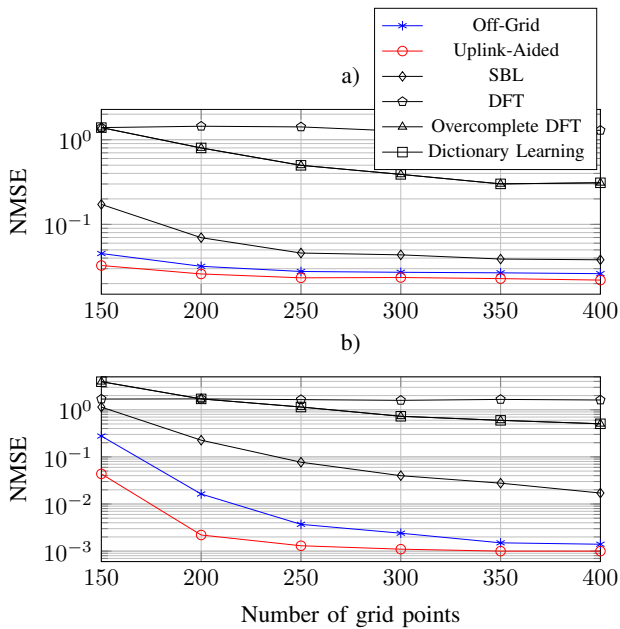


\begin{figure}
\center
\begin{tikzpicture}[scale=0.9]
\begin{semilogyaxis}[
width=5cm,
ylabel={NMSE},grid=both, title={a)},
legend style={at={(0.77,1.57),font=\footnotesize},
anchor=north,legend columns=1},xmin=50,xmax=120,ymin=0.002,ymax=3.49,x=3]
\addplot[mark=asterisk,blue]  coordinates{
  (    50 ,       0.4871  )
  (    60 ,       0.2646  )
  (    70 ,       0.1329  )
  (    80 ,       0.0542  )
  (    90 ,       0.0269  )
  (   100 ,       0.0091  )
  (   110 ,       0.0046  )
  (   120 ,       0.0023  )
};
\addplot[mark=diamond]  coordinates{
  (    50 ,        0.6431 )
  (    60 ,        0.4418 )
  (    70 ,        0.3374 )
  (    80 ,        0.2524 )
  (    90 ,        0.1683 )
  (   100 ,        0.1276 )
  (   110 ,        0.0925 )
  (   120 ,        0.0616 )
};
\addplot[mark=pentagon] coordinates{
  (    50 ,      2.3285)
  (    60 ,      2.3145)
  (    70 ,      2.2090)
  (    80 ,      2.2654)
  (    90 ,      2.4615)
  (   100 ,      2.2285)
  (   110 ,      1.8605)
  (   120 ,      1.2579)
};
\addplot[mark=triangle] coordinates{
  (    50 ,      2.0751)
  (    60 ,      2.1445)
  (    70 ,      2.1299)
  (    80 ,      2.0347)
  (    90 ,      1.8418)
  (   100 ,      1.7594)
  (   110 ,      1.2185)
  (   120 ,      0.7689)
}; \legend{Off-Grid, SBL , DFT, Overcomplete DFT}
\end{semilogyaxis}
\end{tikzpicture}
\begin{tikzpicture}[scale=0.9]
\begin{semilogyaxis}[xlabel={ Number of training pilot symbols},
width=5cm,
ylabel={NMSE},
grid=both, title={b)},
legend style={at={(0.539,1.55),font=\footnotesize},
anchor=north,legend columns=1},xmin=50,xmax=120,ymin=0.0006,ymax=3.49,x=3]
\addplot[mark=asterisk,blue]  coordinates{
  (    40 ,      0.6629)
  (    50 ,      0.4758)
  (    60 ,      0.2740)
  (    70 ,      0.1225)
  (    80 ,      0.0477)
  (    90 ,      0.0167)
  (   100 ,      0.0046)
  (   110 ,      0.0015)
  (   120 ,      0.0007)
  };
\addplot[mark=diamond]  coordinates{
  (    40 ,       0.7438)
  (    50 ,       0.6019)
  (    60 ,       0.4166)
  (    70 ,       0.2995)
  (    80 ,       0.2120)
  (    90 ,       0.1451)
  (   100 ,       0.1086)
  (   110 ,       0.0797)
  (   120 ,       0.0510)
 };
\addplot[mark=pentagon] coordinates{
  (    40 ,      2.0168)
  (    50 ,      2.2128)
  (    60 ,      2.2627)
  (    70 ,      2.2552)
  (    80 ,      2.3032)
  (    90 ,      2.2748)
  (   100 ,      1.8535)
  (   110 ,      1.8213)
  (   120 ,      1.2156)
};
\addplot[mark=triangle] coordinates{
  (    40 ,      1.7895)
  (    50 ,      2.0945)
  (    60 ,      2.1599)
  (    70 ,      2.1376)
  (    80 ,      1.9757)
  (    90 ,      1.6994)
  (   100 ,      1.4353)
  (   110 ,      0.9753)
  (   120 ,      0.6326)
};
\end{semilogyaxis}
\end{tikzpicture}
\caption{NMSE of dowlink channel estimate versus the number of  training pilot symbols for 2D array. a) $\hat L=250$ and SNR $=$ $0$ dB;  b) $\hat L=300$  and SNR $=$ $10$ dB.}
\label{fig-vsT-2D}
\end{figure}
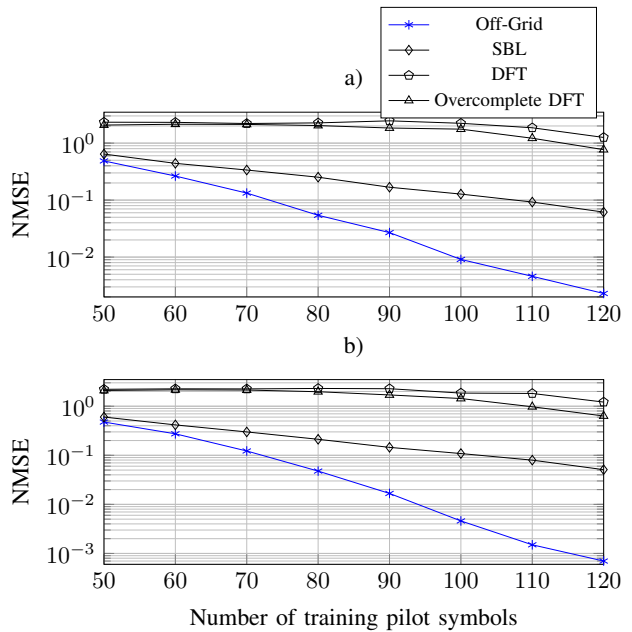


\begin{figure}
\center
\begin{tikzpicture}[scale=0.9]
\begin{semilogyaxis}[
width=5cm,
ylabel={NMSE},grid=both, title={a)},
legend style={at={(0.77,1.57),font=\footnotesize},
anchor=north,legend columns=1},xmin=150,xmax=400,ymin=0.016,ymax=4.2,x=0.839 ]
\addplot[mark=asterisk,blue]  coordinates{
  (   150,       0.2141)
  (   200,       0.1007)
  (   250,       0.0534)
  (   300,       0.0448)
  (   350,       0.0357)
  (   400,       0.0323)
};
\addplot[mark=diamond]  coordinates{
 (   150,        1.1461)
 (   200,        0.3407)
 (   250,        0.2399)
 (   300,        0.2042)
 (   350,        0.1918)
 (   400,        0.1907)
};
\addplot[mark=pentagon] coordinates{
  (   150,          2.2604)
  (   200,          2.5487)
  (   250,          2.6216)
  (   300,          2.5083)
  (   350,          2.2916)
  (   400,          2.4595)
};
\addplot[mark=triangle] coordinates{
  (   150,          4.2826)
  (   200,          2.5487)
  (   250,          2.1827)
  (   300,          1.9688)
  (   350,          1.7031)
  (   400,          1.4910)
};
\legend{Off-Grid, SBL , DFT, Overcomplete DFT}
\end{semilogyaxis}
\end{tikzpicture}
~~
\begin{tikzpicture}[scale=0.9]
\begin{semilogyaxis}[xlabel={Number  of  grid points},
width=5cm,
ylabel={NMSE},grid=both, title={b)},
legend style={at={(0.769,1.45),font=\footnotesize},
anchor=north,legend columns=1},xmin=150,xmax=400,ymin=0.0015,ymax=4.2,x=0.839 ]
\addplot[mark=asterisk,blue]  coordinates{
  (   150,       0.1283)
  (   200,       0.0305)
  (   250,       0.0109)
  (   300,       0.0045)
  (   350,       0.0036)
  (   400,       0.0028)
};
\addplot[mark=diamond]  coordinates{
  (   150,     1.3440)
  (   200,     0.2138)
  (   250,     0.1303)
  (   300,     0.1058)
  (   350,     0.0983)
  (   400,     0.0986)
};
\addplot[mark=pentagon] coordinates{
  (   150,        2.0797)
  (   200,        2.0701)
  (   250,        2.0217)
  (   300,        2.0491)
  (   350,        2.0834)
  (   400,        2.1159)
};
\addplot[mark=triangle] coordinates{
  (   150,      3.7436)
  (   200,      2.0701)
  (   250,      1.5775)
  (   300,      1.4550)
  (   350,      1.1571)
  (   400,      1.0024)
};
\end{semilogyaxis}
\end{tikzpicture}
\caption{NMSE of downlink channel estimate versus the number of  grid points for 2D array with SNR $=$ $0$ dB. a) $T= 80$; b) $T= 100$.} 
 \label{figNg-2D}
\end{figure}
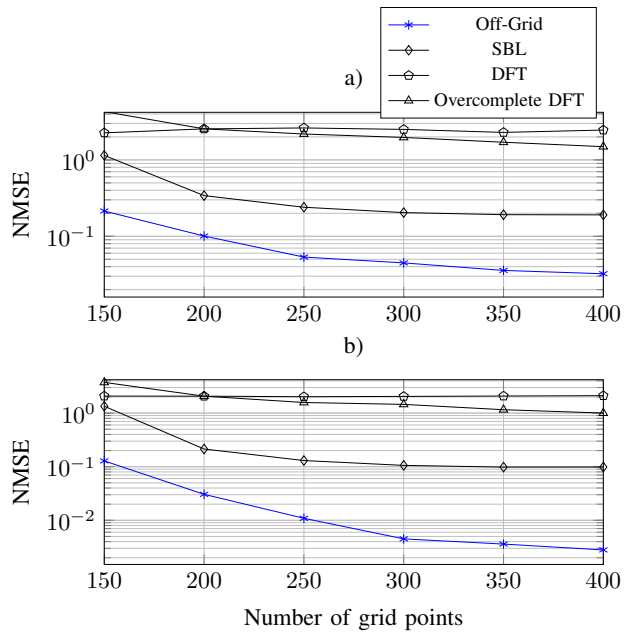

\subsection{Channel Estimation Performance Versus $\hat{L}$ for ULA}

In Fig.~\ref{figNg1}, we study the impact of the number of grid points on the downlink channel estimation performance for ULA. We consider the same scenario as in Section~VI-B, except that the number of  training pilot symbols is fixed to $70$,
and the scattering clusters range from $-90^\circ$ to $90^\circ$.
All the results are obtained by averaging over 200 Monte Carlo channel realizations.
Fig.~\ref{figNg1} shows the NMSE performance of the downlink channel estimate achieved
by the different channel estimation strategies versus  the number of grid points $\hat L$.
It is shown that the overcomplete DFT method and  dictionary learning method achieve the same performance, because there is no benefit in learning the true AoDs which range from $-90^\circ$ to $90^\circ$. The NMSEs of the DFT method, overcomplete DFT method and SBL method coincide with each other at $\hat L=150$ in Fig.~\ref{figNg1}-a and $\hat L=200$ in Fig.~\ref{figNg1}-b, respectively,  because they use the same grid in the case of $N=\hat L$.
The NMSEs of  our methods decrease as the number of grid points increases, and they always outperform the others, no matter what  number of grid points is used.

\subsection{Channel Estimation Performance with 2D Array}
In Figs.~\ref{fig-vsT-2D} and \ref{figNg-2D}, Monte Carlo trials are carried out to investigate the channel estimation performance with the 2D array. Assume that the 2D planar array at the BS is equipped with $20\times 10$ antennas, where
both the horizontal and vertical inter-antenna spacings are a half wavelength. Every  channel realization consists of $N_c=20$ random scattering clusters, and each cluster contains $N_s=20$ subpaths. The AoDs are randomly generated by the 3GPP 3D channel model, where the azimuth AoDs range from $-180^\circ$ to $180^\circ$ and the elevation AoDs range from $-90^\circ$ to $90^\circ$.
The training pilots are randomly generated, the SNR is chosen as $0$ dB or $10$ dB, and $\rho$ in (\ref{equpv3-fix-2D}) is set to $0.95$.
All the results are obtained by averaging over $200$ Monte Carlo channel realizations.
Fig.~\ref{fig-vsT-2D} shows the NMSE performance of the downlink channel estimate achieved by the different channel estimation strategies versus the number of training pilot symbols $T$, and
Fig.~\ref{figNg-2D} shows the NMSE performance of the downlink channel estimate achieved
by the different channel estimation strategies versus  the number of grid points $\hat L$.
It can be seen that 1) the DFT method and overcomplete DFT method give very poor performance, because they can not work for non-ULAs;
2) the standard SBL method outperforms the DFT-based methods, but the performance improvement is not significant;
and 3) our proposed off-grid method indeed works for the 2D array, and it can substantially improve the channel estimation performance.

\section{Conclusion}

The problem  of downlink channel estimation in FDD massive MIMO systems is addressed
in this paper. We provide a novel off-grid model for massive MIMO
channel sparse representation, which can greatly improve
the sparsity and accuracy of the channel representation.  To the best of our knowledge, our
work is the first to utilize an off-grid channel model
to combat  modeling error for channel estimation.
The proposed off-grid model and the SBL-based
framework have wide applicability. They do not require any prior
knowledge about the sparsity of channels, nor the
variance of noises, and all the parameters are automatically tuned by the in-exact MM algorithm. Extending the results
to MUs with multiple  antennas is
straightforward in the framework of SBL.

\section*{Appendix}

\appendices

\subsection{Proof of Lemma~1}

The non-decreasing property can be achieved  as
\begin{align}
&\ln p(\y,  \alpha^{(i+1)},\bm{\gamma}^{(i+1)},\bm\beta^{(i+1)} )  \notag\\
\ge& \mathcal{U}( \alpha^{(i+1)},\bm{\gamma}^{(i+1)},\bm\beta^{(i+1)}  | \alpha^{(i+1)},\bm{\gamma}^{(i+1)},\bm\beta^{(i)} )\label{eqi1}\\
\ge& \mathcal{U}( \alpha^{(i+1)},\bm{\gamma}^{(i+1)},\bm\beta^{(i)}  | \alpha^{(i+1)},\bm{\gamma}^{(i+1)},\bm\beta^{(i)} )\label{eqi2}\\
=& \ln p(\y,  \alpha^{(i+1)},\bm{\gamma}^{(i+1)},\bm\beta^{(i)} ) \label{eqi3}\\
\ge& \mathcal{U}( \alpha^{(i+1)},\bm{\gamma}^{(i+1)},\bm\beta^{(i)}  | \alpha^{(i+1)},\bm{\gamma}^{(i)},\bm\beta^{(i)} )\label{eqi4}\\
\ge& \mathcal{U}( \alpha^{(i+1)},\bm{\gamma}^{(i)},\bm\beta^{(i)}  | \alpha^{(i+1)},\bm{\gamma}^{(i)},\bm\beta^{(i)} )\label{eqi5}\\
=& \ln p(\y,  \alpha^{(i+1)},\bm{\gamma}^{(i)},\bm\beta^{(i)} ) \label{eqi6}\\
\ge& \mathcal{U}( \alpha^{(i+1)},\bm{\gamma}^{(i)},\bm\beta^{(i)}  | \alpha^{(i)},\bm{\gamma}^{(i)},\bm\beta^{(i)} )\label{eqi7}\\
\ge& \mathcal{U}( \alpha^{(i)},\bm{\gamma}^{(i)},\bm\beta^{(i)}  | \alpha^{(i)},\bm{\gamma}^{(i)},\bm\beta^{(i)} )\label{eqi8}\\
=& \ln p(\y,  \alpha^{(i)},\bm{\gamma}^{(i)},\bm\beta^{(i)} ), \label{eqi9}
\end{align}
where (\ref{eqi1}),  (\ref{eqi4}) and  (\ref{eqi7}) follow (\ref{eqMM1});  (\ref{eqi3}),  (\ref{eqi6}) and  (\ref{eqi9}) follow (\ref{eqMM2}); and  (\ref{eqi2}), (\ref{eqi5}) and (\ref{eqi8}) follow (\ref{eqM3}), (\ref{eqM2}) and (\ref{eqM1}), respectively.

\subsection{Proof of Lemma~2}

Letting  $q(\w)$ be an arbitrary distribution, the lower bound of $\ln p(\y, \alpha,\bm{\gamma}, \bm\beta )$ can be written as
\begin{align}
\ln p(\y, \alpha,\bm{\gamma}, \bm\beta ) =&  \ln \int p(\w, \y, \alpha,\bm{\gamma},\bm\beta)  d \w \notag\\
=&  \ln \int  q(\w ) \frac{ p(\w, \y, \alpha,\bm{\gamma},\bm\beta) }{ q(\w) }  d \w \notag\\
\ge&  \int  q(\w) \ln \frac{ p(\w, \y, \alpha,\bm{\gamma},\bm\beta) }{ q(\w) }  d \w, \label{eqApex1}
\end{align}
where Jensen's inequality is applied in the last step. The equality holds when
$ \frac{ p(\w, \y, \alpha,\bm{\gamma},\bm\beta) }{ q(\w) } =c$
for a constant $c$ that does not depend on $\w$. As $q(\w)$ is a distribution, we have
$\int q(\w) d \w  =1$.
This further indicates that
\begin{align}
c=\int p(\w, \y, \alpha,\bm{\gamma},\bm\beta)  d\w = p( \y, \alpha,\bm{\gamma},\bm\beta)
\end{align}
and
\begin{align}
q(\w)
= p(\w| \y, \alpha,\bm{\gamma},\bm\beta ).\label{eqApex2}
\end{align}
With (\ref{eqApex1}) and (\ref{eqApex2}), it is easy to check that the constructed surrogate function $\mathcal{U}(\alpha,\bm{\gamma}, \bm\beta| \dot{\alpha},\dot{\bm{\gamma}},\dot{\bm\beta})$ always satisfies (\ref{eqMM1}) and (\ref{eqMM2}) for  any fixed $(\dot{\alpha},\dot{\bm{\gamma}},\dot{\bm\beta})$.

To prove (\ref{eqMM3}), we first rewrite the left side of (\ref{eqMM3}) as
\begin{align}
  &\left.\frac{\partial \mathcal{U}(\alpha,\dot{\bm{\gamma}}, \dot{\bm\beta}| \dot{\alpha},\dot{\bm{\gamma}},\dot{\bm\beta})}{\partial \alpha}\right|_{\alpha=\dot{\alpha}}\notag\\
=&\left.\int p(\w| \y, \dot{\alpha},\dot{\bm{\gamma}},\dot{\bm\beta} ) \frac{\partial \ln p(\w, \y, \alpha,\dot{\bm{\gamma}},\dot{\bm\beta)} }{ \partial \alpha  }  d \w\right|_{\alpha=\dot{\alpha}}\notag\\
=&\left.\int  \frac {p(\w| \y, \dot{\alpha},\dot{\bm{\gamma}},\dot{\bm\beta} )} { p(\w, \y, \alpha,\dot{\bm{\gamma}},\dot{\bm\beta}) }
\frac{\partial p(\w, \y, \alpha,\dot{\bm{\gamma}},\dot{\bm\beta}) }{ \partial \alpha  }  d \w\right|_{\alpha=\dot{\alpha}}\notag\\
=& \frac {1} { p(\y, \dot{\alpha},\dot{\bm{\gamma}},\dot{\bm\beta}) }  \left.\int\frac{\partial p(\w, \y, \alpha,\dot{\bm{\gamma}},\dot{\bm\beta}) }{ \partial \alpha  }  d \w\right|_{\alpha=\dot{\alpha}}\notag\\
=& \frac {1} { p(\y, \dot{\alpha},\dot{\bm{\gamma}},\dot{\bm\beta}) }  \cdot
\left.\frac{\partial p( \y, \alpha,\dot{\bm{\gamma}},\dot{\bm\beta}) }{ \partial \alpha  }\right|_{\alpha=\dot{\alpha}}.\label{eqpg1}
\end{align}
On the other hand, the right side of (\ref{eqMM3}) is
\begin{align}
&\frac{\partial  \ln p(\y,  \alpha, \dot{\bm{\gamma}},\dot{\bm\beta} ) }{\partial \alpha}
= \frac {1} { p(\y, {\alpha}_d,\dot{\bm{\gamma}},\dot{\bm\beta}) }
\frac{\partial p( \y, \alpha,\dot{\bm{\gamma}},\dot{\bm\beta}) }{ \partial \alpha  }.\label{eqpg2}
\end{align}
Combining (\ref{eqpg1}) and (\ref{eqpg2}), we achieve the equality in (\ref{eqMM3}).  Since the proofs for  (\ref{eqMM4})--(\ref{eqMM5}) can be similarly achieved, they are omitted for brevity.

\subsection{Proof of Lemma~3}

For $\alpha$, ignoring the independent terms, the objective function  in (\ref{eqM1}) can be rewritten as
\begin{align}
&\mathcal{U}(\alpha,\bm{\gamma}^{(i)}, \bm\beta^{(i)}| \alpha^{(i)},\bm{\gamma}^{(i)},\bm\beta^{(i)} )\notag\\
=& \int p(\w| \y, \alpha^{(i)},\bm{\gamma}^{(i)},\bm\beta^{(i)}  ) \ln  p(\y | \w, \alpha, \bm\beta^{(i)})
d \w\notag\\
&~~~~~~+ \int p(\w| \y, \alpha^{(i)},\bm{\gamma}^{(i)},\bm\beta^{(i)}  ) \ln  p(\alpha)
d \w  \notag\\
=& -\alpha \int p(\w| \y, \alpha^{(i)},\bm{\gamma}^{(i)},\bm\beta^{(i)}  ) \left\| \y- \bm\Phi(\bm\beta^{(i)}) \w \right\|_2^2
d \w  \notag\\
&~~~~~~~~~~~~~~~~~~~ + T\ln \alpha   +(a)\ln\alpha -b\alpha  \notag\\
=& (T+a)\ln\alpha  -\alpha  (b + \eta(\alpha^{(i)},\bm{\gamma}^{(i)},\bm\beta^{(i)} ) ). \label{eq-upal}
\end{align}
Since (\ref{eq-upal}) is a strick concave function  related to $\alpha$, setting its derivative to zero gives  the unique optimal solution
\begin{align*}
\alpha^{(i+1)} = \frac{T + a}{ b + \eta(\alpha^{(i)},\bm{\gamma}^{(i)},\bm\beta^{(i)} ) }.
\end{align*}

\subsection{Proof of Lemma~4}
For $\bm\gamma$, ignoring the independent terms of the objective function  in (\ref{eqM2}), we obtain
\begin{align*}
&\mathcal{U}(\alpha^{(i+1)},\bm{\gamma}, \bm\beta^{(i)}| \alpha^{(i+1)},\bm{\gamma}^{(i)},\bm\beta^{(i)} )\notag\\
=&\int p(\w| \y, \alpha^{(i+1)},\bm{\gamma}^{(i)},\bm\beta^{(i)}  ) \ln   p(\w|\bm{\gamma}) d \w\notag\\
&~~~~~~~~~~~~ +\int p(\w| \y, \alpha^{(i+1)},\bm{\gamma}^{(i)},\bm\beta^{(i)}  ) \ln   p(\bm{\gamma}) d \w  \notag\\
=& - \ln |\diag(\bm\gamma^{-1})|  +(a)\sum_{l=1}^{\hat L} \ln\gamma_i -  b \sum_{l=1}^{\hat L} \gamma_i  \notag\\
-&\int p(\w| \y, \alpha^{(i+1)},\bm{\gamma}^{(i)},\bm\beta^{(i)}  ) \left( (\w)^H\diag(\bm\gamma) \w \right)  d \w\notag\\
=& \sum_{l=1}^{\hat L} \ln \gamma_l  +(a)\sum_{l=1}^{\hat L} \ln\gamma_i -  b \sum_{l=1}^{\hat L} \gamma_i  \notag\\
&~~~~~~~~~-  \tr\left( \bm\Xi(\alpha^{(i+1)},\bm{\gamma}^{(i)},\bm\beta^{(i)})\cdot \diag(\bm\gamma) \right).
\end{align*}

Differentiating w.r.t.  each $\gamma_l$ yields
\begin{align*}
&\frac{\partial\mathcal{U}(\alpha^{(i+1)},\bm{\gamma}, \bm\beta^{(i)}| \alpha^{(i+1)},\bm{\gamma}^{(i)},\bm\beta^{(i)} ) }{\partial \gamma_l}\notag\\
=&\frac{a+1}{\gamma_l} - b - \left[\bm\Xi(\alpha^{(i+1)},\bm{\gamma}^{(i)},\bm\beta^{(i)})\right]_{ll}.
\end{align*}
Then, setting the derivative to zero and solving for  $\gamma_l$  give the unique optimal solution
\begin{align*}
\gamma_l^{(i+1)}= \frac{a+1}{b+  \left[\bm\Xi(\alpha^{(i+1)},\bm{\gamma}^{(i)},\bm\beta^{(i)})\right]_{ll} }.
\end{align*}

\subsection{Derivation for Eq. (\ref{eqderbeta})}\label{Apbetadev}
Ignoring the independent terms, the objective function in (\ref{eqM3}) becomes
\begin{align}
&\mathcal{U}(\alpha^{(i+1)},\bm{\gamma}^{(i+1)}, \bm\beta| \alpha^{(i+1)},\bm{\gamma}^{(i+1)},\bm\beta^{(i)} )\notag\\
=&\int p(\w| \y, \alpha^{(i+1)},\bm{\gamma}^{(i+1)},\bm\beta^{(i)}  )\ln p(\y | \w, \alpha^{(i+1)}, \bm\beta) d \w \notag\\
=&-\alpha^{(i+1)} \int p(\w| \y, \alpha^{(i+1)},\bm{\gamma}^{(i+1)},\bm\beta^{(i)}  )\notag\\
 &~~~~~~~~~~~~~~~~~~~~~~~~~~~~~~~~~~~~~~~~~\cdot\left\| \y  -  \bm\Phi(\bm\beta) \w  \right\|_2^2 d \w \notag\\
=&  -\alpha^{(i+1)} \left\| \y  -  \bm\Phi(\bm\beta) \bm\mu(\alpha^{(i+1)},\bm{\gamma}^{(i+1)},\bm\beta^{(i)}) \right\|_2^2\notag\\
&~~~~~~~~~- \alpha^{(i+1)}  \tr\left(\bm\Phi(\bm\beta)\bm\Sigma(\alpha^{(i+1)},\bm{\gamma}^{(i+1)},\bm\beta^{(i)}) \bm\Phi^H(\bm\beta)\right). \notag
\end{align}
For ease of notation, we simply denote $\bm\mu(\alpha^{(i+1)},\bm{\gamma}^{(i+1)},\bm\beta^{(i)})$ and $ \bm\Sigma(\alpha^{(i+1)},\bm{\gamma}^{(i+1)},\bm\beta^{(i)}) $ by $\bm\mu^{(i)}$ and $\bm\Sigma^{(i)}$, respectively.
Calculating the derivative of each term in  the above equality  w.r.t.  $\beta_l$, we obtain
\begin{align*}
&\frac{\partial \left\|\y- \bm\Phi(\bm\beta) \bm\mu^{(i)} \right\|_2^2}{\partial \beta_l }\\
=&\frac{\partial\left\|\y^{(i)}_{-l} - \mu_{l}^{(i)} \cdot\X (\a (\hat\vartheta_{ l}+\beta_{ l}) )  \right\|_2^2}{\partial \beta_l } \notag\\
=& 2 \mathrm{Re}\left( (  \a' (\hat\vartheta_{ l}+\beta_{ l}))^H \X^H\X \a (\hat\vartheta_{ l}+ \beta_{ l})\right) \cdot  |\mu_{l}^{(i)}|^2\notag\\
 &~~~~~~~~   - 2 \mathrm{Re}\left(  (\a' (\hat\vartheta_{ l}+ \beta_{ l}))^H \X^H \cdot (\mu_{l}^{(i)})^* \y_{-l}^{(i)}\right)
\end{align*}
and
\begin{align*}
&\frac{\partial \tr\left(\bm\Phi(\bm\beta)\bm\Sigma^{(i)}\bm\Phi^H(\bm\beta)\right) }{\partial \beta_l }\notag\\
=&2 \mathrm{Re}\left(   (\a' (\hat\vartheta_{ l}+\beta_{ l}))^H \X^H\X \a (\hat\vartheta_{ l}+ \beta_{ l}) \right) \cdot \chi_{ll}^{(i)}\notag\\
&+ 2 \mathrm{Re}\left(  (\a' (\hat\vartheta_{ l}+ \beta_{ l}))^H \X^H \X \cdot\sum_{j\ne l} \chi_{jl}^{(i)} \a (\hat\vartheta_{j}+ \beta_{ j}) \right).
\end{align*}
Hence, the derivative of $\mathcal{U}(\alpha^{(i+1)},\bm{\gamma}^{(i+1)}, \bm\beta| \alpha^{(i+1)},\bm{\gamma}^{(i+1)},\bm\beta^{(i)} )$ w.r.t $\beta_l$ is same as (\ref{eqderbeta1}).

\subsection{Proof of Theorem~5}
According to Theorem~2-b in \cite{razaviyayn2014successive}, the block MM algorithm will converge to a stationary solution if the following additional conditions are satisfied:
\begin{itemize}
  \item All the properties in  (\ref{eqMM1})--(\ref{eqMM5}) hold true with the  surrogate function.

  \item At least two of the problems (\ref{eqM1})--(\ref{eqM3}) have a unique solution.
\end{itemize}
Lemmas~2--4 guarantee that the above two conditions hold true, respectively

\bibliographystyle{IEEEtran}
\bibliography{offgridCE}

\end{document}